\documentclass[useAMS,usenatbib]{mn2e}
\usepackage{epsfig}
\usepackage{float}
\usepackage{placeins}
\usepackage{graphicx}
\usepackage{epsfig}
\usepackage{bm}% bold math
\usepackage{amsfonts}
\usepackage{amssymb}
\usepackage{times}
\usepackage{natbib}
\usepackage{color}

\title[Flux and spectral variations in BL Lacerate]{Multi-band optical variability studies of BL Lacertae}
\author[Agarwal et al.]
{Aditi Agarwal$^{1,2}$\thanks{E-mail: aditi@aries.res.in} and
Alok C.\ Gupta$^{1,2}$
\\ 
$^{1}$Aryabhatta Research Institute of Observational Sciences (ARIES),
Manora Peak, Nainital -- 263002, India\\
$^{2}$Department of Physics, DDU Gorakhpur University, Gorakhpur - 273009, India \\
}
\begin{document}
\newdimen\digitwidth
\setbox0=\hbox{2}
\digitwidth=\wd0
\catcode `#=\active
\def#{\kern\digitwidth}

\date{Accepted ....... Received  ......; in original form ......}

\pagerange{\pageref{firstpage}--\pageref{lastpage}} \pubyear{2014}

\maketitle

\label{firstpage}

\begin{abstract}
We monitored BL Lacertae for 13 nights in optical B, V, R, and I bands during October and November 2014 including quasi-simultaneous observations
in V and R bands using two optical telescopes
in India. We have studied multi-band optical flux variations, colour variation and spectral changes in this blazar. Source was found to be active during
the whole monitoring period and showed significant intraday variability on 3 nights in V and R filters while displayed hints of variability on
6 other dates in R passband and on 2 nights in V filter.
From the colour-magnitude analysis of the source we found that the spectra of the target gets flatter as it becomes brighter on intra-night timescale.
Using discrete correlation technique, we found that intraday light curves in both V and R filters are almost consistent and well correlated
with each other. We also generated spectral energy distribution (SED) of the target using the B, V, R, and I data sets for all 13 nights which
could help us investigate the physical process responsible for the observed variations in BL Lacertae objects.
We also discuss possible physical causes of the observed spectral variability.

 \end{abstract}
 
\begin{keywords}
BL Lacertae objects: general --- : galaxies --- active --- quasars: individual -- BL Lacertae
\end{keywords}

\section{Introduction}
\label{sec:Introduction}

BL Lacertae objects and high polarization sources called as flat spectrum radio quasars (FSRQs) together constitute a violently variable class of
Active Galactic Nuclei (AGNs) known as blazars (e.g., Blandford \& Rees 1978, Ghisellini et al.\ 1997). Common blazar characteristics include high polarization, synchrotron
emission from the relativistic jet, rapid flux variability, core-dominated radio morphology, flat radio spectrum and many more
owing to the relativistic motion of plasma in the jets extending on even Mpc scales, pointing at angles $\leq$ 10$^{\circ}$ with the line of sight (LOS)
thus causing the observed emission to be relativistically beamed (e.g., Urry \& Padovani 1995).
Blazars are known to be variable at all accessible time scales over entire electro-magnetic spectra. Magnitude changes of few hundredth
to tenths over a time scale of a day is called intraday variability (IDV) or micro-variability (Wagner \& Witzel 1995, Rector \& Perlman 2003)
which help us in probing innermost regions of AGNs; changes from several days to few months are usually known as
short time variability (STV); while those taking from several months to many years are usually called long term variability (LTV; Gupta et al. 2004),
in last two classes blazar variability can usually exceed even $\sim$ 5 mag.

Flux variability usually leads to spectral changes due to the changes in the spectra of emitting electrons along with the activity
associated with the relativistic Doppler boosted jets. The spectral energy distribution (SED) of blazars display two well-defined broad
spectral components (Mukherjee et al. 1997; Weekes 2003). Based on the location of these peaks, we have low energy
peaked blazars (LBLs) whose first component peaks in near-IR (NIR)/optical while the second component usually peaks at GeV energies and high
energy peaked blazars (HBLs) where first peak is in UV/X-rays while the second peak at TeV energies (Padovani \& Giommi 1995, Abdo et al.\
2010). During flaring state the peak is found
to be shifted towards higher frequencies. 
Spectral properties were linked with the source luminosity by Fossati et al.\ (1997) such that HBLs were the low luminosity objects with high space
density and high electron density thus high magnetic field while the objects with high luminosity, low space density plus low magnetic field
were named LBLs. Ratio of X-ray flux in the 0.3-3.5 kev band to the radio flux density at 5GHz is also used to classify as
low-synchrotron-peaked or high-synchrotron-peaked blazars (LSPs or HSPs) based
on whether the ratio is smaller or greater than 10$^{-11.5}$ respectively (Padovani \& Giommi 1996). In addition to LSPs and HSPs we also have
intermediate synchrotron peaked blazars (ISPs) also for which SED peaks are located at intermediate frequencies (Sambruna, Maraschi \& Urry 1996).
One can also define LSPs as those blazars with energy peak frequency of their synchrotron hump, $\nu_{\rm peak}$ $\sim$ 10$^{13-14}$~Hz, ISPs as those with $\nu_{\rm peak}$
$\sim$ 10$^{15-16}$~Hz while those with $\nu_{\rm peak}$ $\sim$ 10$^{17-18}$~Hz as HSPs. The above classification was proposed by Nieppola,
Tornikoski \& Voltaoja (2006) based on the studies of 300 BL Lacertae objects. From the observational point of view it has been found that LBLs are more
optically variable than HBLs (Stocke et al.\ 1985; 1989; Heidt \& Wagner 1996)

BL Lacertae ($\alpha_{2000.0}$ = 22h 02m 43.29s
$\delta_{2000.0}$ = $+42^{\circ} 16^{'} 39.98^{\prime \prime}$) located at a redshift value of z = 0.069 (Miller \& Hawley 1977)
with a moderately bright elliptical host galaxy (Wurtz, Stocke \& Yee 1996), is a prototype of the blazar class of AGNs
which has been intensively studied since its discovery by Schmitt (1968) with radio source VRO 42.22.01 and characterized by a featureless
spectrum.
BL Lacertae is classified as a low-frequency peaked blazar (Fossati et al.\ 1998; Abdo et al.\ 2010) with first component of its SED peaking in
near-IR (NIR)/optical region which can be explained by synchrotron emission from the helical jet (Raiteri et al.\ 2009),
thus making optical studies of the BL Lacertae helpful to understand them in greater detail. Optical observations
can also indicate the possible presence of other components in addition to synchrotron continuum.
There is good evidence for the presence of IDV and colour variability in some BL Lacertae objects and our
source is among those (e.g., Carini et al.\ 1992, Clements \& Carini 2001, Gu et al.\ 2006, H.E.S.S.~Collaboration et al.\ 2011).
Due to its pronounced variability over entire electromagnetic spectra it has been continuously monitored with Whole Earth Blazar
Telescope (WEBT)\footnote{http://www.oato.inaf.it/blazars/webt/} since 1997 till today; which involves four major WEBT
multi-frequency campaigns, two of them were organized in 1999 supplemented with ASCA and Beppo SAX observations (Ravasio
et al.\ 2002) while the other two provided densely sampled, high precision, long term campaigns carried out during May 2000 -
January 2001
and May 2001 - February 2002 (Villata et al.\ 2002; B{\"o}ttcher et al.\ 2003; Villata et al.\ 2004a,b).
\begin{table*}
\begin{tabular}{c  c }
\begin{tabular}{lccll} \\
\multicolumn{2}{|c|}{Table 1: } \\
\multicolumn{2}{|c|}{A: 1.04 m telescope, ARIES, Nainital, India} \\
\multicolumn{2}{|c|}{B: 1.3 m Ritchey-Chretien Cassegrain optical}\\
\multicolumn{2}{|c|}{  telescope, ARIES, Nainital, India}\\\hline
Site:                 &A                              &B                    \\\hline
Telescope:          &1.04-m RC Cassegrain           &1.30m Ritchey-Chr\'etien  \\
CCD model:          & Wright 2K CCD                 &Andor 2K CCD             \\
Chip size:          & $2048\times2048$ pixels       &$2048\times2048$ pixels  \\
Pixel size:         &$24\times24$ $\mu$m            &  $13.5\times13.5$ $\mu$m \\
Scale:              &0.37\arcsec/pixel              &       0.535\arcsec/pixel \\
Field:              & $13\arcmin\times13\arcmin$    & $18\arcmin\times18\arcmin$  \\
Gain:               &10 $e^-$/ADU                   & 1.4 $e^-$/ADU               \\
Read Out Noise:     &5.3 $e^-$ rms                  & 4.1 $e^-$ rms               \\
Binning used:       &$2\times2$                     &    $2\times2$               \\
Typical seeing :    & 1\arcsec to 2.8\arcsec        &   1.2\arcsec to 2.0\arcsec   \\\hline
\end{tabular}

\begin{tabular}{lcll} \\
\multicolumn{2}{|c|}{Table 2}\\\hline

 Date of        & Telescope      & Number of data points \\
observations    &     & Filters     \\
(yyyy mm dd)     & & (B,V,R,I)      \\\hline
 
 2014 10 26    & A  &1,42,42,1  \\
  2014 10 27   & A &1,42,43,1  \\
  2014 11 04  & A  &1,31,31,1 \\
 2014 11 05   & A &1,44,44,1 \\
2014 11 06    & A &1,42,42,1 \\
 2014 11 07   & A  &1,35,35,1 \\
 2014 11 08   & A  &1,45,45,1 \\
  2014 11 09  & A  &1,49,49,1 \\
  2014 11 13  & A   &1,1,1,1 \\
 2014 11 15  & B    &1,1,1,1 \\
  2014 11 16 & B  &1,46,46,1 \\
 2014 11 21  & B  &1,1,1,1 \\
  2014 11 22 & B   &1,52,52,1  \\ \hline
\end{tabular}

\end{tabular}
\caption{Details of telescopes and instruments}
\vspace{-0.15in}\caption{Observation log of photometric observations of BL Lacertae}
\end{table*}

\begin{figure}
\epsfig{figure=  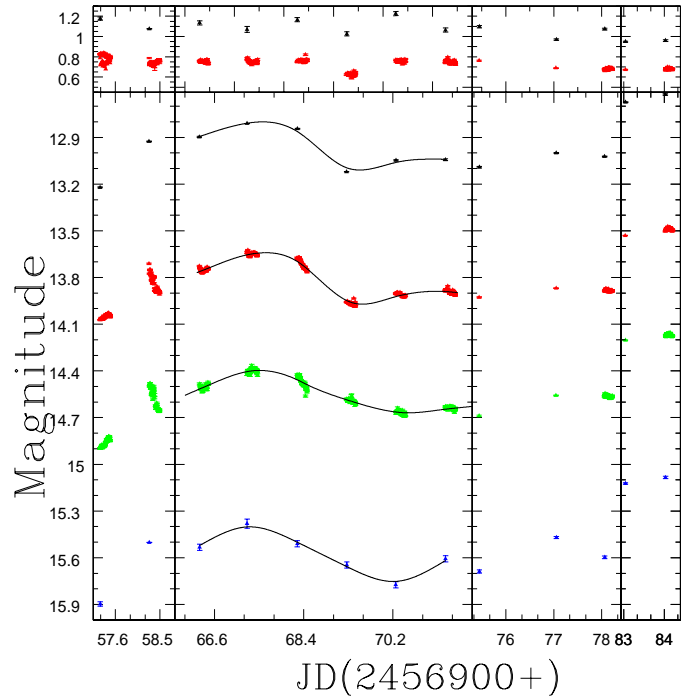,height=3.7in,width=3.6in,angle=0}
\caption{Light curves for BL Lacertae covering full monitoring period; blue denotes B filter; green denotes V filter;  red, R filter; and black, I filter.
Y axis is the magnitude while x-axis is Julian Date (JD).
In the top panel red colour represents (V-R) colour variation while black is for (B-I) $-$ 1.7 colour. Offset of 1.7 is used with (B-I) plot
to avoid its eclipsing with (V-R) plot.
Solid line represents a cubic spline interpolation during the continuous 6 nights
monitoring in B, V, R, and I filters.}
\label{LC_BL}
\end{figure}

\begin{figure*}
\epsfig{figure=  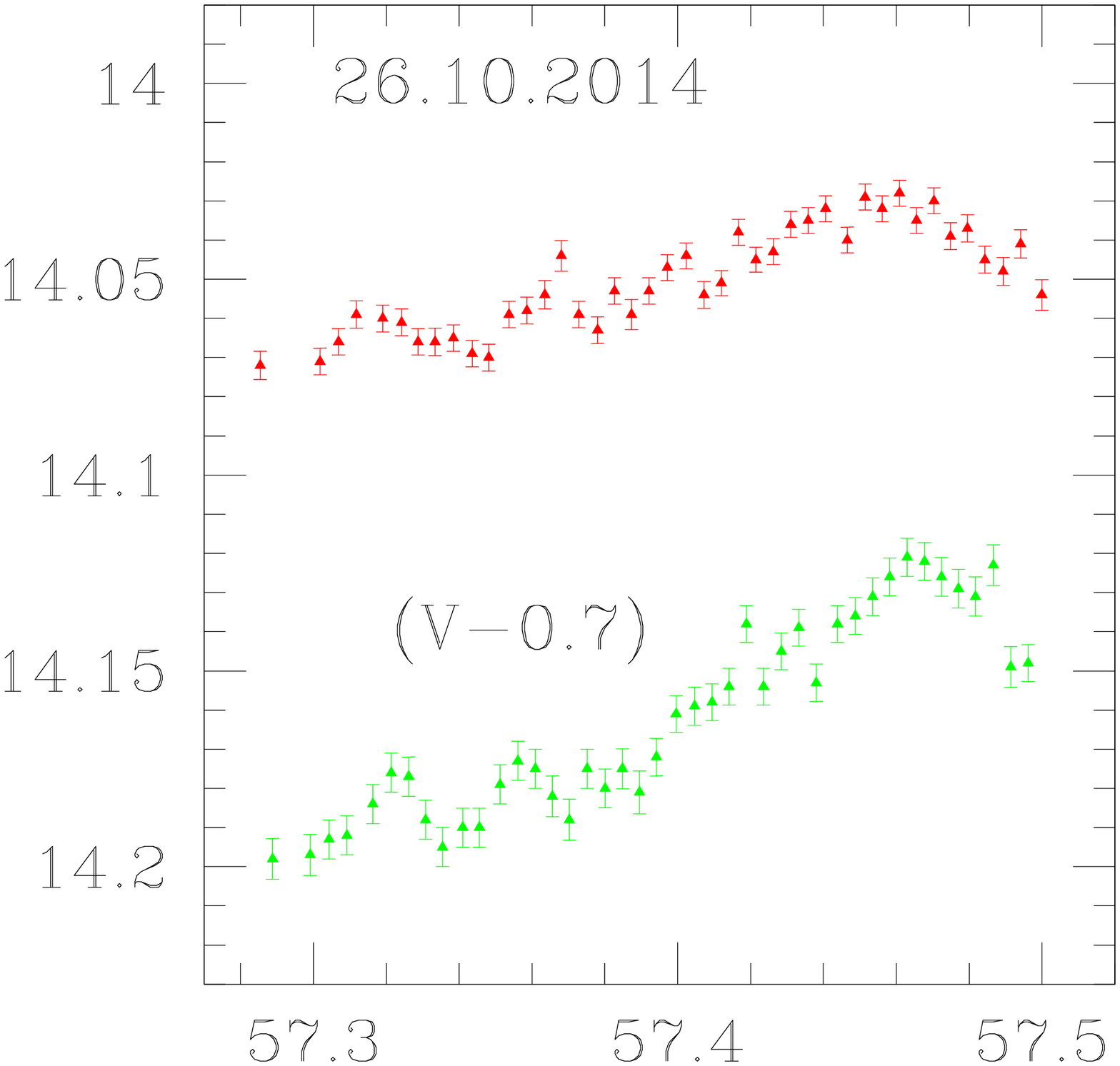,height=1.567in,width=1.59in,angle=0}
 \epsfig{figure=  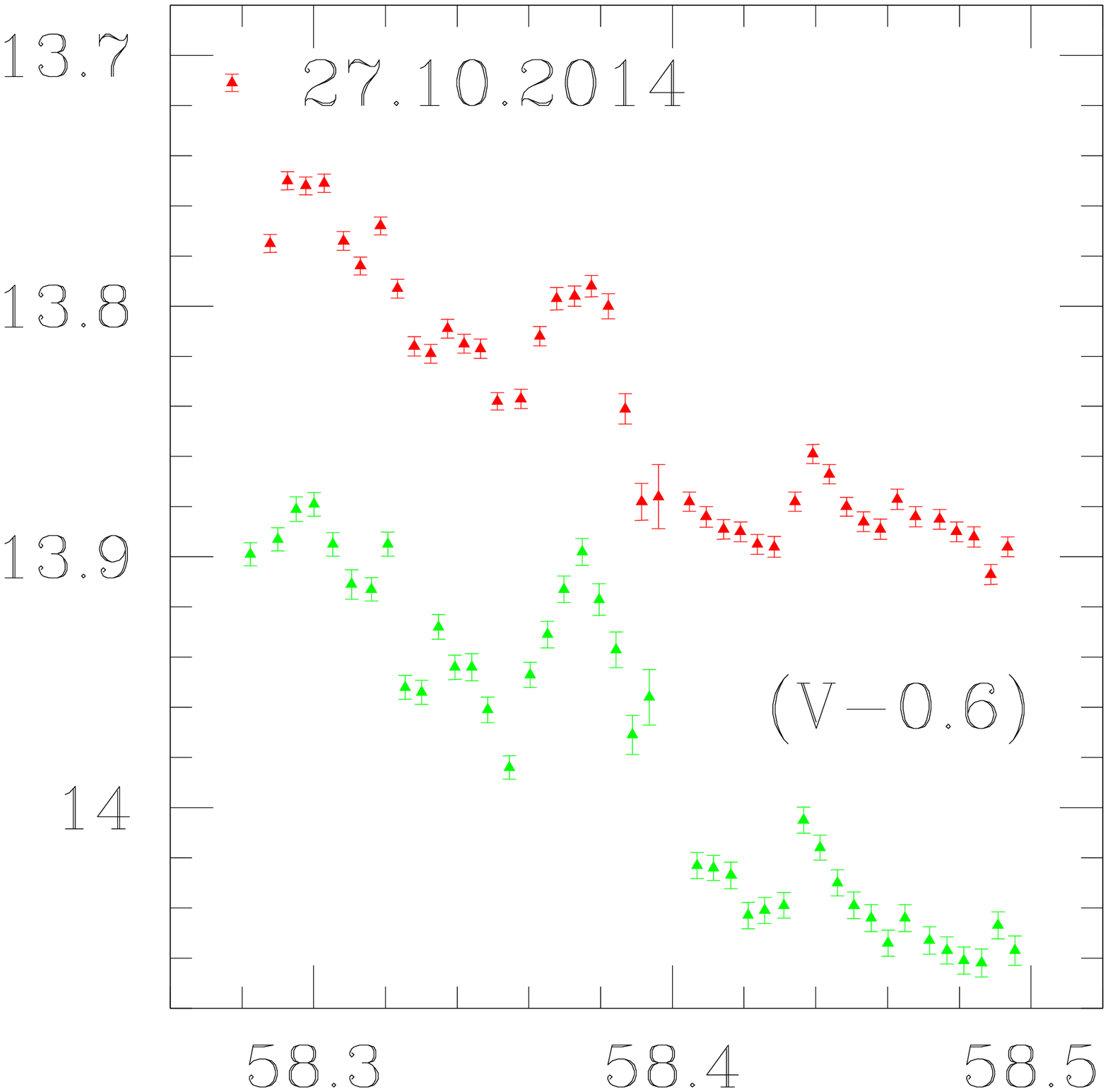,height=1.567in,width=1.59in,angle=0}
\epsfig{figure= 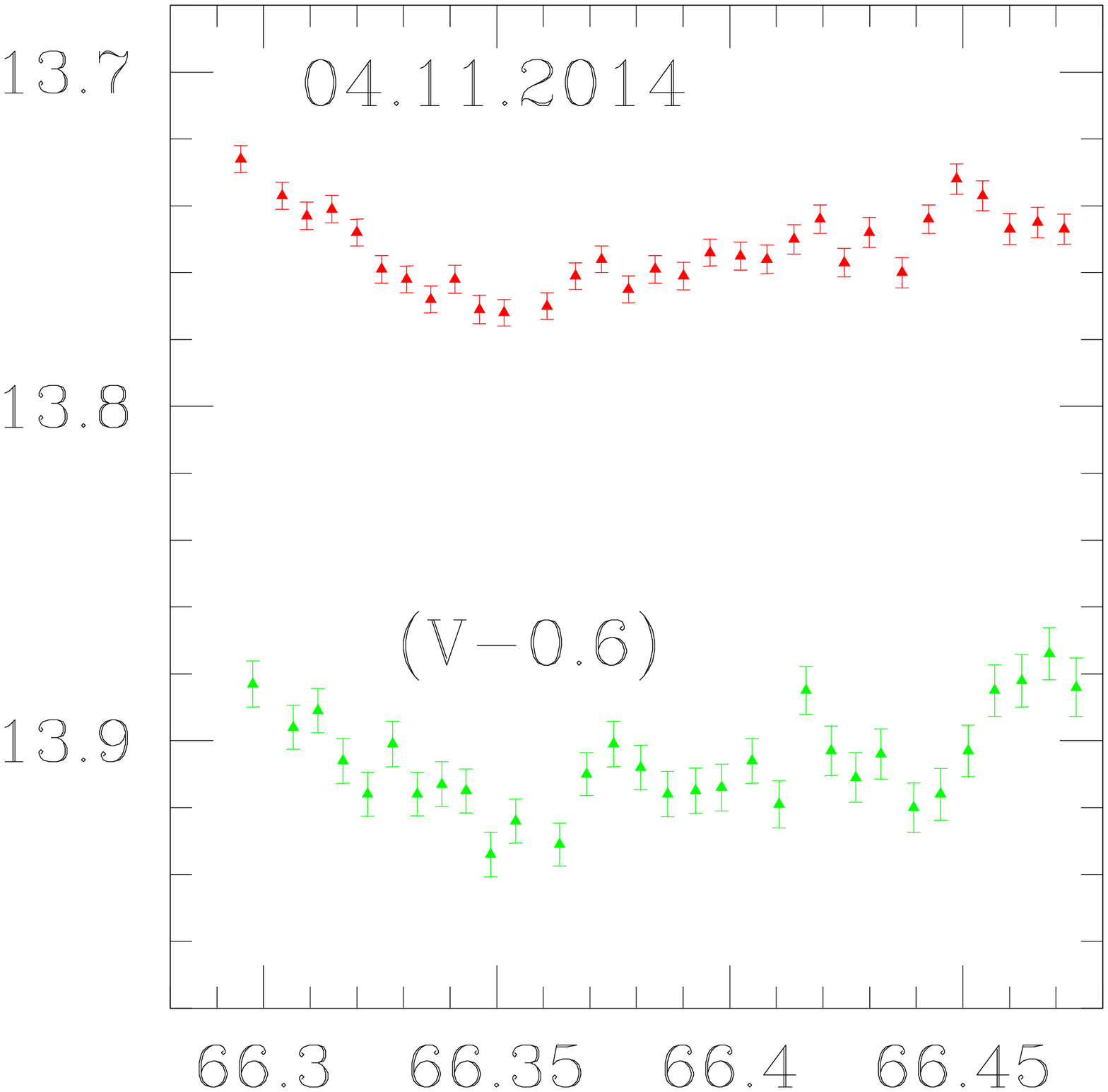 ,height=1.567in,width=1.59in,angle=0}
\epsfig{figure=  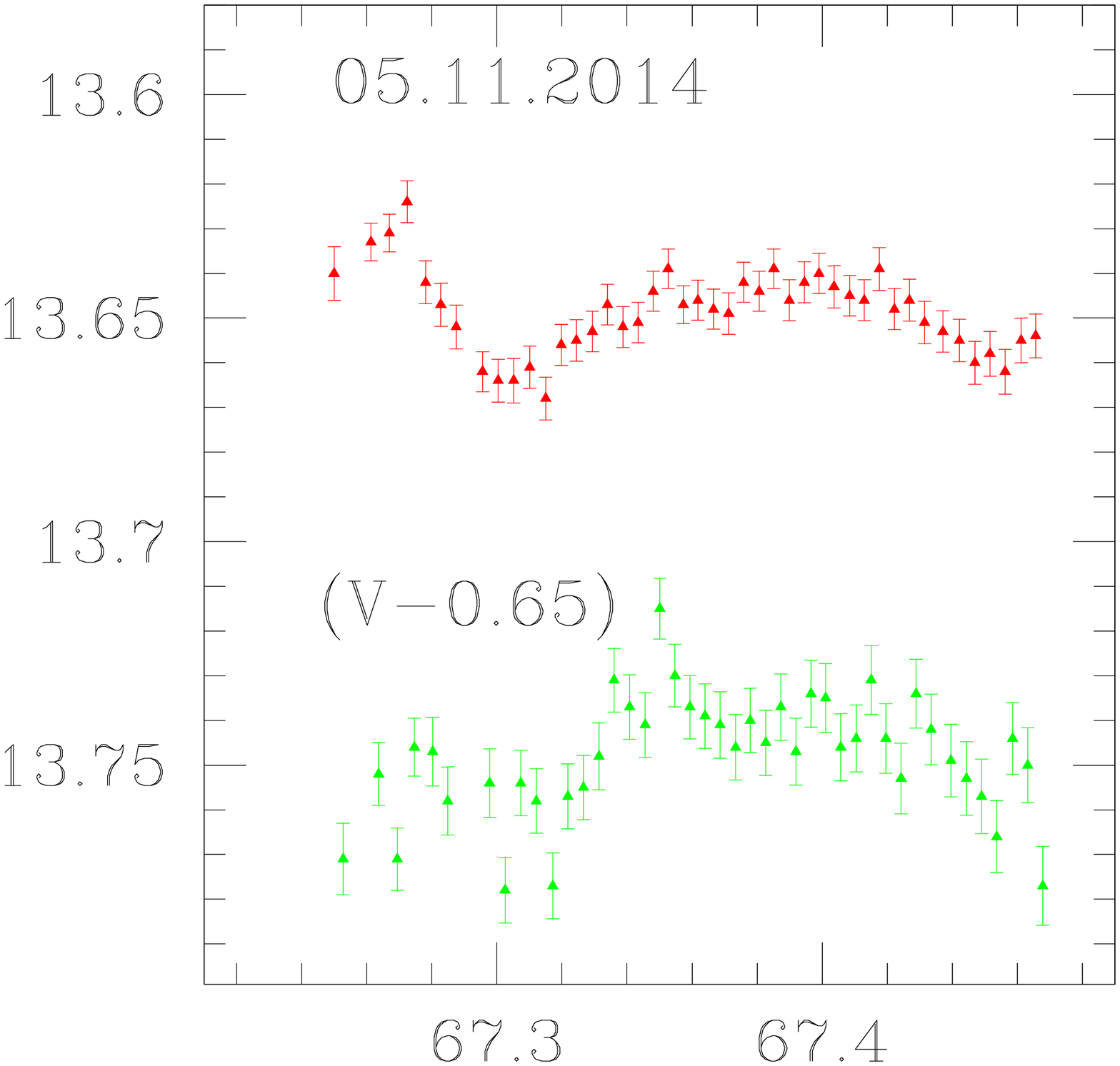,height=1.567in,width=1.59in,angle=0}
\epsfig{figure=  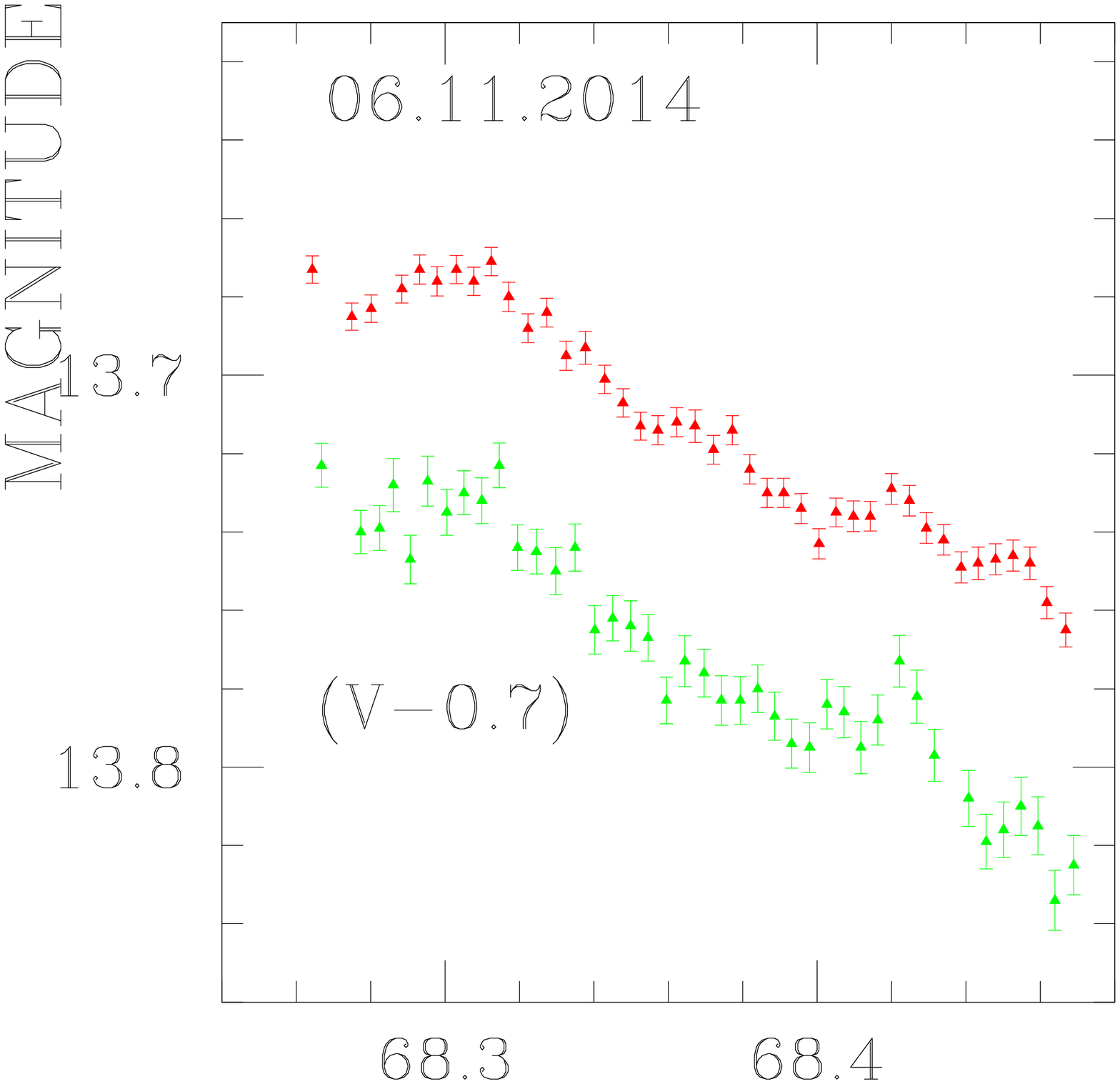,height=1.567in,width=1.59in,angle=0}
\epsfig{figure=  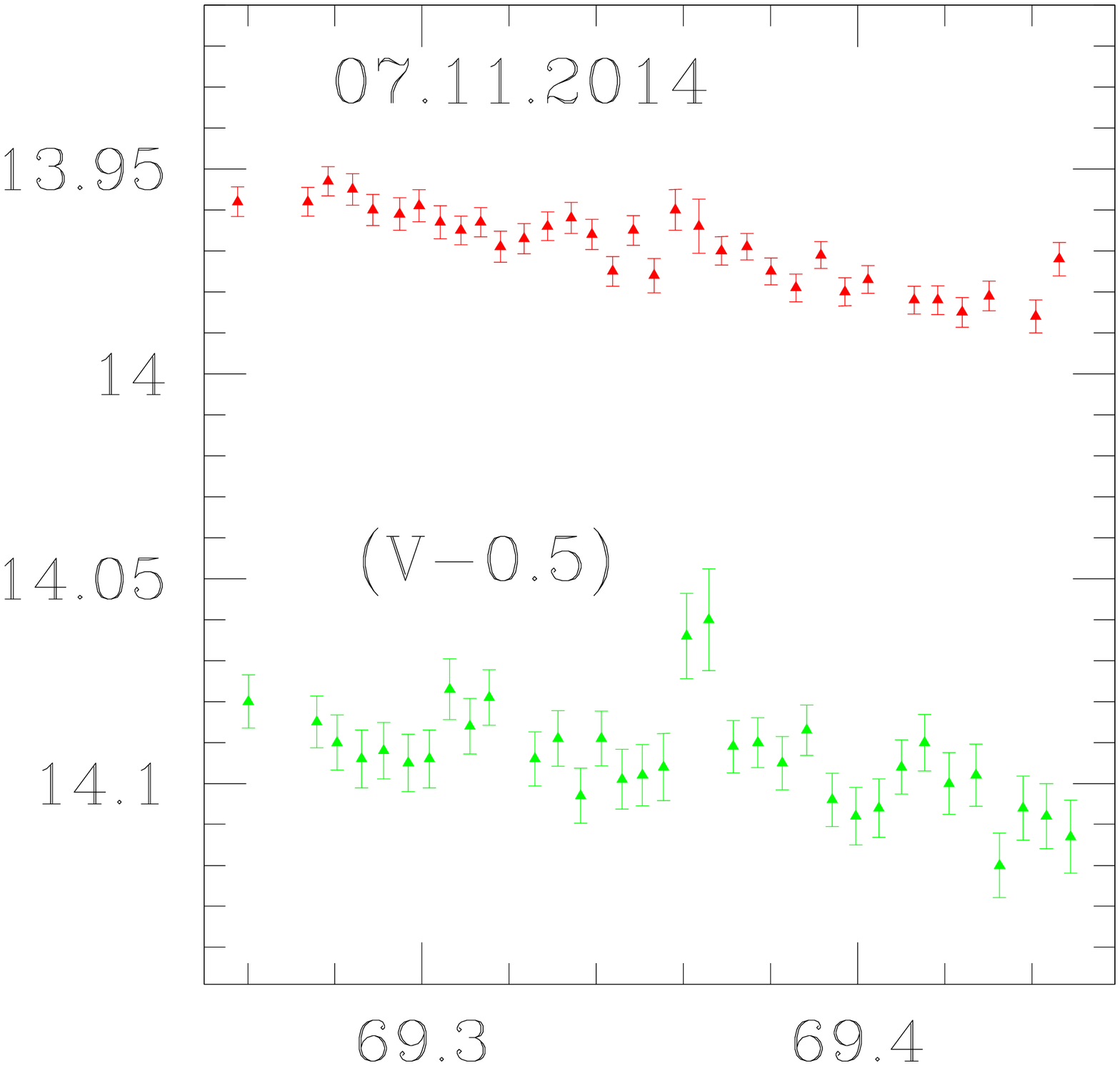,height=1.567in,width=1.59in,angle=0}
\epsfig{figure=  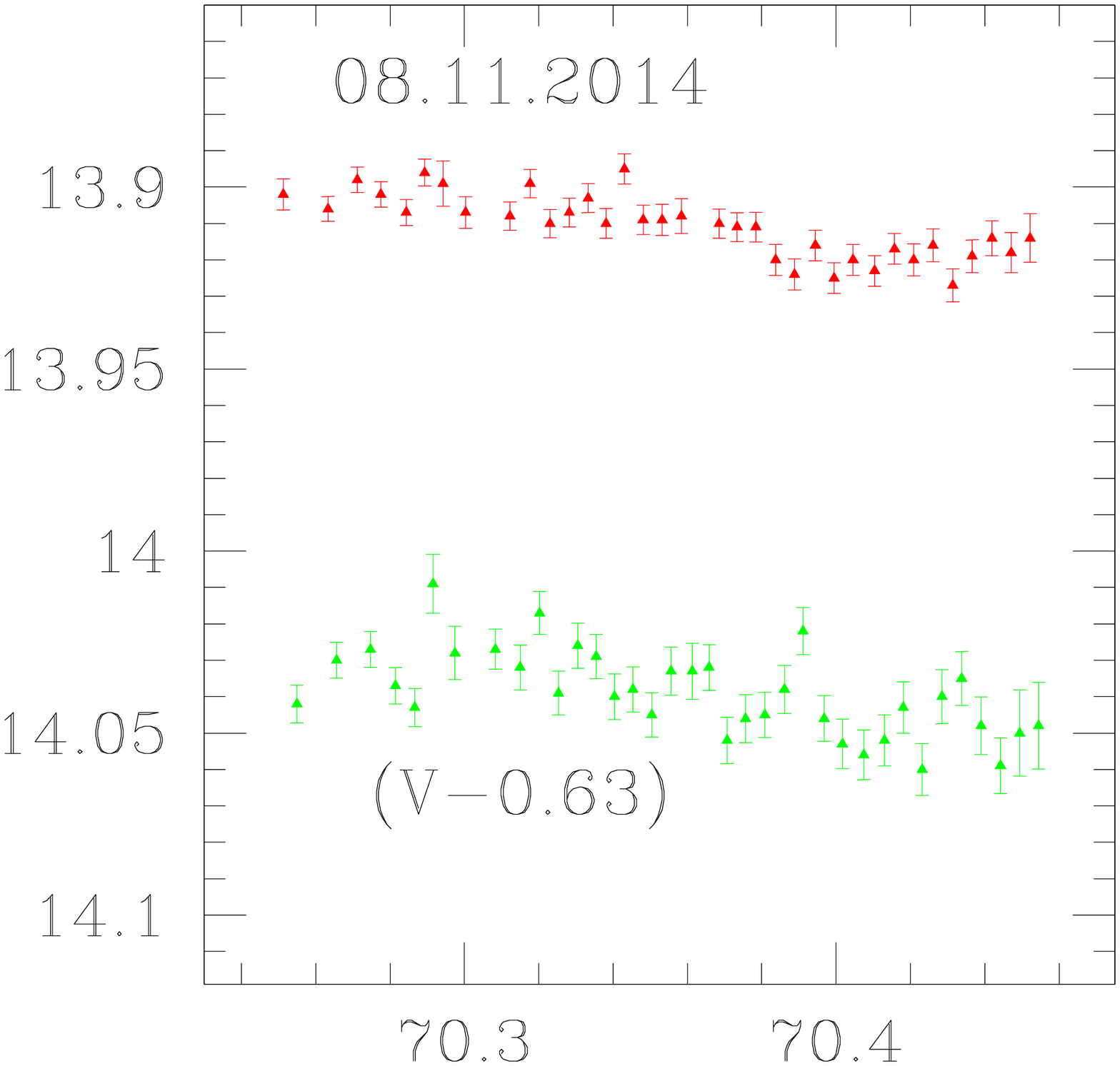,height=1.567in,width=1.59in,angle=0}
\epsfig{figure= 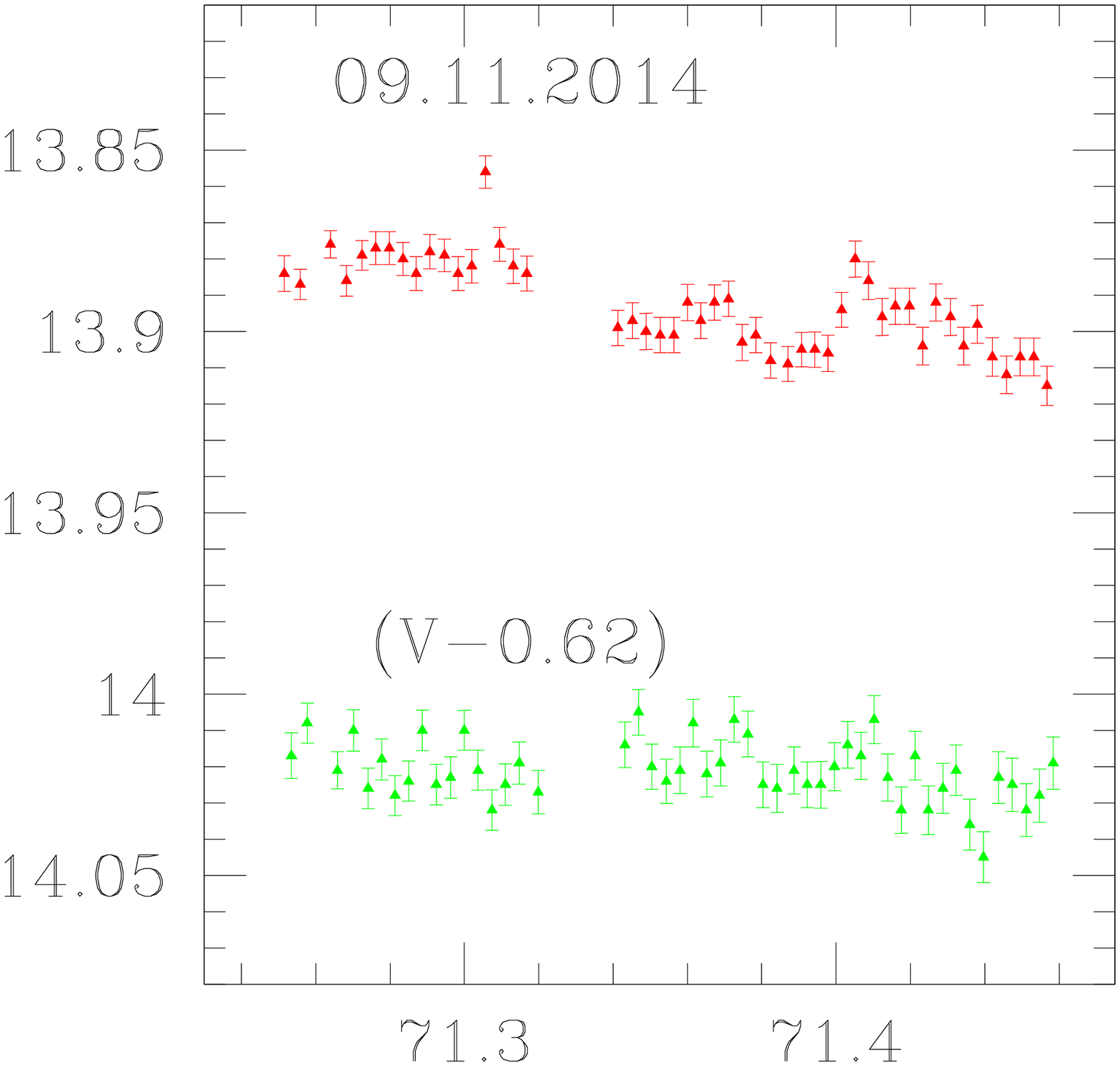,width=1.59in,height=1.567in,angle=0}
 \epsfig{figure= 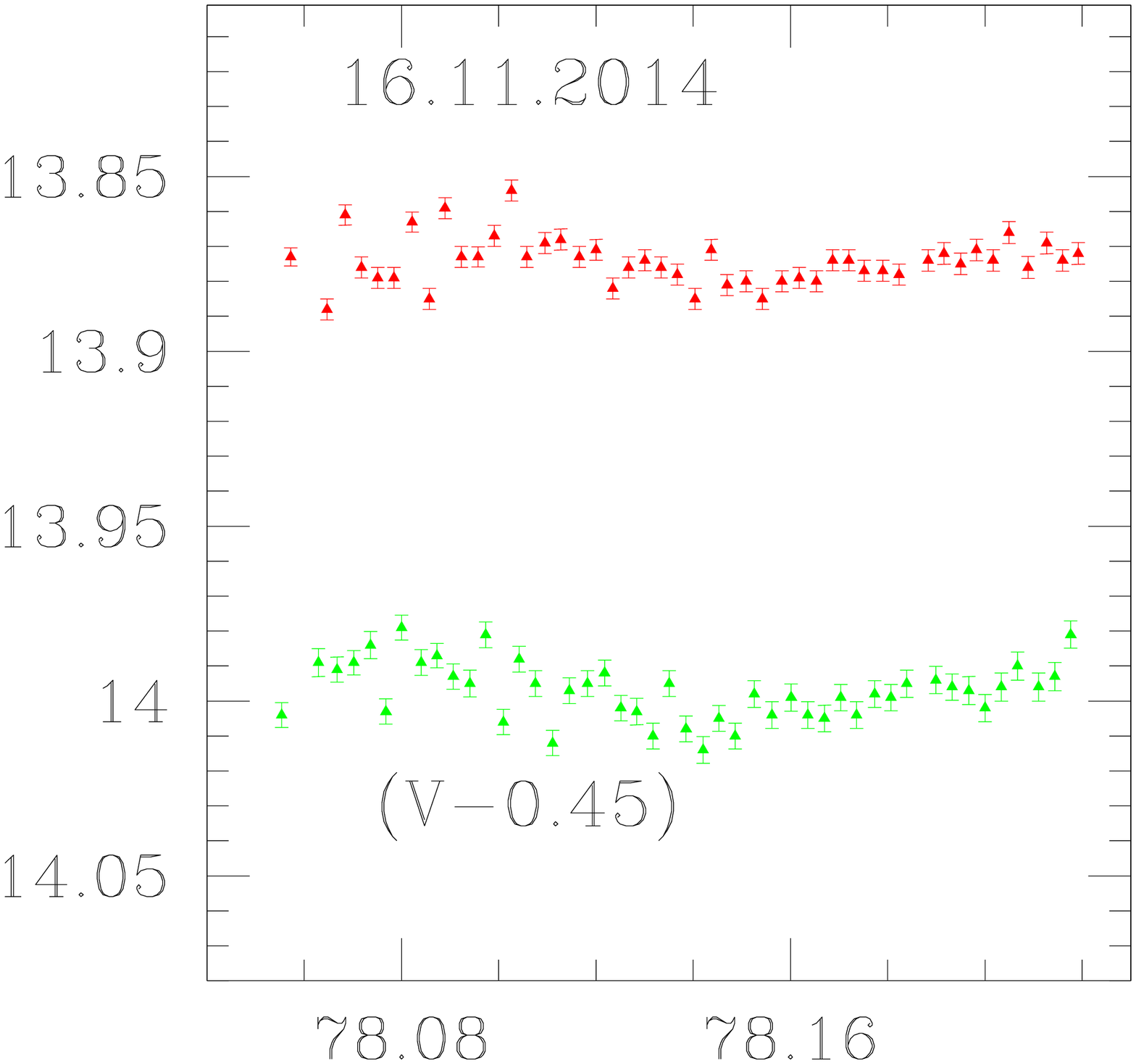,width=1.59in,height=1.567in,angle=0}
  \epsfig{figure= 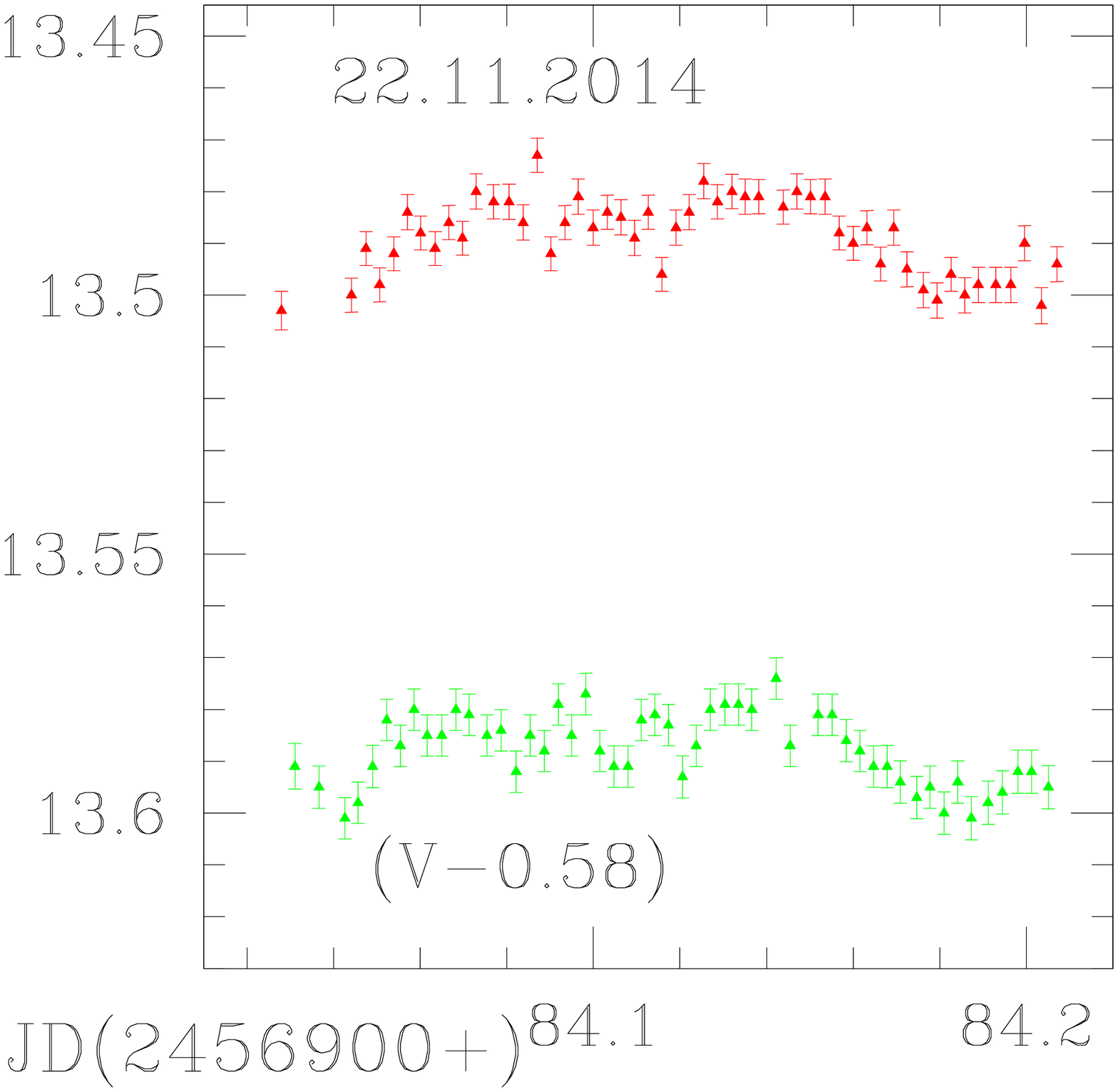,width=1.59in,height=1.567in,angle=0}

  \caption{Light curves for BL Lacertae; green denotes V filter;  red, R filter. Y axis is the magnitude in each plot while X axis is JD.
Observation date is indicated in each plot.}
\label{LC_BL}
\end{figure*}

Carini et al.\ (1992) studied two optically bright sources, OJ 287 and BL Lacertae using a decade long
photometric observations where they found IDV of the order of 0.08 mag/hr in former and 0.01
mag/hr for later. Authors also detected hints of bluer-when-brighter (BWB) trend but any relation
between spectral and flux changes were absent. Later, Villata et al.\ (2002, 2004a) confirmed the
presence of BWB trend in short isolated outburst but not on long timescales for BL Lacertae, indicating
two different components are operating. Heidt \& Wagner (1996) found 80\% of complete BL Lacerate
objects to show IDV confirming it to be their intrinsic nature.
During the 1997 optical outburst of our target, strong correlation was found between V-R
colour and R magnitude with the correlation coefficient reaching $\sim$ 0.7 in 11 nights.
Although blazar colour variability has been extensively studied since many years, but is still a hot topic of discussion, particularly for
BL Lacertae, where underlying bright elliptical galaxy contaminating the photometric magnitudes mainly at low flux levels, could be one of the
reasons. Rise and decay timescales for the target have also been studied by Papadakis et al.\ (2003) and found them increasing with frequency.
Significant flux-flux correlation (Raiteri et al.\ 2013) has been established from extensive optical
studies of the BL Lacertae objects.

In this paper, our aim is to have a flavor of variability nature of the source BL Lacertae on shortest possible timescale;
study colour and spectral change; and  to have an idea about the process occurring in
the vicinity of the central super massive black hole (SMBH) through photometric studies of flux variations.

The structure of the paper is as follows: In Section 2, we present the details of the observations and data reduction procedures.
Section 3 describes various analysis
techniques used, while the variability results are reported in Section 4, followed by correlation studies in Section 5 and
colour variability studies in Section 6.
Section 7 describes the spectral changes in the source and finally discussion and conclusion are given in Section 8.

\hspace*{-0.5in}
\begin{table*}
\caption{ Results of IDV observations of BL Lacertae. Column 1 is the date of observation, column 2 indicates the band in which observations were taken,
column 3 represents number of data points in a particular band, in the next three columns i.e. 4, 5, and 6 we have listed the results for C-statistic,
F-test and $\chi^{2}$-test, respectively, followed by variability status in column 7 and finally
the intra-night variability amplitude is given in column 8. }
\textwidth=7.0in
\textheight=10.0in
\vspace*{0.2in}
\noindent

\begin{tabular}{ccccccccc} \hline \nonumber

 Date       & Band   &N    &  $C_{1},C_{2},C_{avg}$       & F-test  &$\chi^{2}$test  &   Variable    &A\% \\
               &            &         &                               &$F_{1},F_{2},F,F_{c}(0.99),F_{c}(0.999)$ &$\chi^{2}_{1},
\chi^{2}_{2},\chi^{2}_{av}, \chi^{2}_{0.99}, \chi^{2}_{0.999}$  & & \\\hline 

 26.10.2014   & V     & 42 & 3.5587, 3.0478, 3.3033 & 12.67, 9.29, 10.98, 2.09, 2.69 & 328.66, 238.62, 283.64, 65.95, 74.75  & Var  & 7.67 \\
              & R     & 42 & 2.8122, 2.8627, 2.8375 & 7.90, 8.19, 8.05, 2.09, 2.69     & 248.54, 249.99, 249.26, 65.95, 74.75  & Var  & 4.37 \\
             &(V-R)   & 42 & 2.1598, 1.6191, 1.8894 & 4.66, 2.62, 3.64, 2.09, 2.69    & 125.14, 69.93, 97.53, 65.95, 74.75    & PV & 6.24 \\  
 27.10.2014   & V     & 40 & 3.3657, 3.9467, 3.6562 & 11.33, 15.58, 13.45, 2.13, 2.76 & 400.44, 504.13, 452.28, 62.43, 72.05 & Var & 22.79 \\
              & R     & 40 & 9.2009, 9.1329, 9.1669 & 84.66, 83.41, 84.03, 2.13, 2.76 & 3006.4, 2791.2, 2898.8, 62.43, 72.05 & Var  & 15.69 \\
              &(V-R)  & 40 & 1.9050, 1.3330, 1.6190 & 3.63, 1.78, 2.70, 2.13, 2.76    & 124.42, 133.04, 128.73, 62.43, 72.05 &  PV  & 6.94\\             
 04.11.2014   & V     & 31 & 1.5193, 1.7475, 1.6334 & 2.31, 3.05, 2.68, 2.39, 3.22 & 151.19, 161.24, 156.21, 50.89, 59.70 & PV  & 5.92 \\
              & R     & 31 & 2.2875, 2.3634, 2.3254 & 5.23, 5.58, 5.41, 2.39, 3.22 & 140.91, 85.86, 113.38, 50.89, 59.70 & PV  & 4.56 \\
              &(V-R)  & 31 & 0.7035, 1.0318, 0.8676 & 0.50, 1.06, 0.78, 2.39, 3.22 & 9.47, 12.96, 11.21, 50.89, 59.70    & NV & -- \\             
 05.11.2014   & V     & 43 & 1.5208, 1.5623, 1.5416 & 2.31, 2.44, 2.38, 2.38, 2.66 & 68.56, 49.08, 58.82, 66.21, 76.08 & PV  & 6.21 \\
              & R     & 43 & 2.1314, 2.0030, 2.0672 & 4.54, 4.01, 4.28, 2.38, 2.66 & 203.64, 156.18, 179.91, 66.21, 76.08 & PV  & 4.35 \\
              &(V-R)  & 43 & 1.5656, 1.5073, 1.5364 & 2.45, 2.27, 2.36, 2.38, 2.66 & 86.54, 58.20, 72.37, 66.21, 76.08 & NV & -- \\              
 06.11.2014   & V     & 41 & 5.4723, 5.1830, 5.3276 & 29.95, 26.86, 28.40, 2.09, 2.69  & 776.44, 463.29, 619.86, 64.95, 74.74  & Var  & 11.06 \\
              & R     & 41 & 6.8721, 6.8382, 6.8552 & 47.22, 46.76, 46.99, 2.09, 2.69 & 1804.10, 1692.90, 1748.50, 64.95, 74.74 & Var  & 9.38 \\
              &(V-R)  & 41 & 1.2996, 1.2186, 1.2591 & 1.69, 1.48, 1.59, 2.09, 2.69 & 51.70, 32.09, 41.89, 64.95, 74.74 & NV & -- \\
 07.11.2014   & V     & 34 & 1.2005, 0.9753, 1.0879 & 1.44, 0.95, 1.20, 2.29, 3.04   & 13.55, 12.25, 12.9, 54.77, 63.87  & NV  & -- \\
              & R     & 35 & 2.0896, 2.1564, 2.1230 & 4.37, 4.65, 4.51, 2.26, 2.98  & 65.64, 104.59, 85.11, 56.06, 65.25  & PV  & 5.17 \\
              &(V-R)  & 32 & 1.2802, 0.9024, 1.0912 & 1.64, 0.81, 1.23, 2.35, 3.15  & 17.34, 11.42, 14.38, 52.19, 61.09 & NV & -- \\              
 08.11.2014   & V     & 36 & 1.5381, 1.5667, 1.5524 & 2.37, 2.45, 2.41, 2.23, 2.93  & 46.53, 31.72, 39.12, 57.34, 66.62   & NV  & -- \\
              & R     & 35 & 1.5678, 1.8589, 1.7134 & 2.46, 3.45, 2.96, 2.26, 2.98  & 71.42, 96.85, 84.13, 56.06, 65.25  & PV  & 3.14 \\
              &(V-R)  & 35 & 1.1850, 1.1754, 1.1802 & 1.40, 1.38, 1.39, 2.26, 2.98  & 31.86, 23.91, 27.88, 56.06, 65.25 & NV & -- \\              
 09.11.2014   & V     & 49 & 0.9388, 1.0879, 1.0134 & 0.88, 1.18, 1.03, 1.98, 2.49  & 24.12, 22.76, 23.44, 73.68, 84.04  & NV  & -- \\
              & R     & 49 & 1.5703, 2.0319, 1.8011 & 2.47, 4.13, 3.30, 1.98, 2.49  & 91.42, 100.00, 95.71, 73.68, 84.04 & PV  & 5.86 \\
              &(V-R)  & 48 & 0.9525, 1.4293, 1.1909 & 0.91, 2.04, 1.47, 1.99, 2.51 & 27.96, 44.81, 36.38, 72.44, 82.72 & NV & \\ 
 16.11.2014   & V     & 46 & 1.0017, 0.8398, 0.9207 & 1.00, 0.70, 0.85, 2.02, 2.57 & 41.30, 75.73, 58.51, 69.96, 80.08 & NV  & -- \\
              & R     & 46 & 1.2413, 1.1424, 1.1919 & 1.54, 1.30, 1.42, 2.02, 2.57 & 35.02, 45.83, 40.42, 69.96, 80.08 & NV  & -- \\
              &(V-R)  & 46 & 0.9653, 0.6830, 0.8242 & 0.93, 0.47, 0.70, 2.02, 2.57 & 38.53, 50.76, 44.64, 69.96, 80.08 & NV & -- \\               
 22.11.2014   & V     & 51 & 1.1749, 0.9398, 1.0573 & 1.38, 0.88, 1.13, 1.95, 2.44 & 69.79, 50.85, 60.32, 76.15, 86.66 & NV  & -- \\
              & R     & 51 & 1.6082, 1.3373, 1.4728 & 2.59, 1.79, 2.19, 1.95, 2.44 & 134.92, 95.32, 115.12, 76.15, 86.66 & PV  & 2.96 \\
              &(V-R)  & 51 & 1.1525, 0.8490, 1.0007 & 1.33, 0.72, 1.02, 1.95, 2.44 & 66.15, 40.32, 53.23, 76.15, 86.66 & NV & -- \\                
\hline
\end{tabular}
\noindent
Var : Variable, PV : probable variable, NV : Non-Variable     \\
\end{table*}

\section{\bf Observations and Data Reductions}

Observations of the BL Lacertae were carried out using two optical telescopes of India, one is the 1.04 m Sampuranand telescope 
with f/13 Cassegrain focus located at Aryabhatta Research Institute of observational sciencES (ARIES), Nainital and the other one is the
1.3-m Devasthal fast optical telescope (DFOT) of ARIES, Nainital, India. Both these telescopes are equipped with CCD detectors and broadband
Johnson UBV and Cousins RI filters. Technical parameters and instrumental details are summarized in Table 1.
The optical photometric data for BL Lacertae was gathered between 25 Oct to 23 Nov 2014. For the pre-processing of the raw data we used
standard procedures in the
IRAF\footnote{IRAF is distributed by the National Optical Astronomy Observatories, which are operated
by the Association of Universities for Research in Astronomy, Inc., under cooperative agreement with the
National Science Foundation.} software following the steps described below. Bias frames were taken at regular intervals covering whole night
which are used to generate a master bias for that particular observation night by taking median of all bias frames. This master bias was subtracted
from all twilight flat frames and the image frames. Next step in the initial processing is of flat fielding when master flat in each filter is
generated by median combine of all the flat frames in a particular passband. Next, normalized master flat for each passband is generated by
which each source image frame was divided to remove pixel to pixel inhomogenities. Finally, cosmic ray removal was carried out for all
source image frames.
Data processing was then done using the Dominion Astronomical
Observatory Photometry (DAOPHOT II) software (Stetson 1987; Stetson 1992) to get the instrumental magnitudes of the BL Lac and the comparison stars
by performing the concentric circular aperture photometry technique. For every night, we carried out aperture photometry with four different
aperture radii, i.e., $\sim 1 \times$~FWHM, $2 \times$~FWHM, 
$3 \times$~FWHM and $4 \times$~FWHM, out of which aperture radii of $2 \times$~FWHM was finally selected for our analysis as it provided the best
S/N ratio. We observed more than three local standard stars on the same field. Out of these, we used those two comparison stars 
which had magnitude and colour similar to that of the blazar to avoid any error occurring from differences in the photon statistics in the
differential photometry of the source. Comparison stars used for differential photometry of the blazar are stars B and C of the finding chart from
the webpage\footnote{http://www.lsw.uni-heidelberg.de/projects/extragalactic/charts/2200+420.html}. Since the BL Lacertae and the standard stars
magnitudes were obtained simultaneously under same air mass and weather conditions, so the flux values are considered reliable. To get the calibrated
magnitude of the target, that non variable standard star was chosen which had colour closest to the BL Lacertae.

\begin{figure*}
\epsfig{figure=  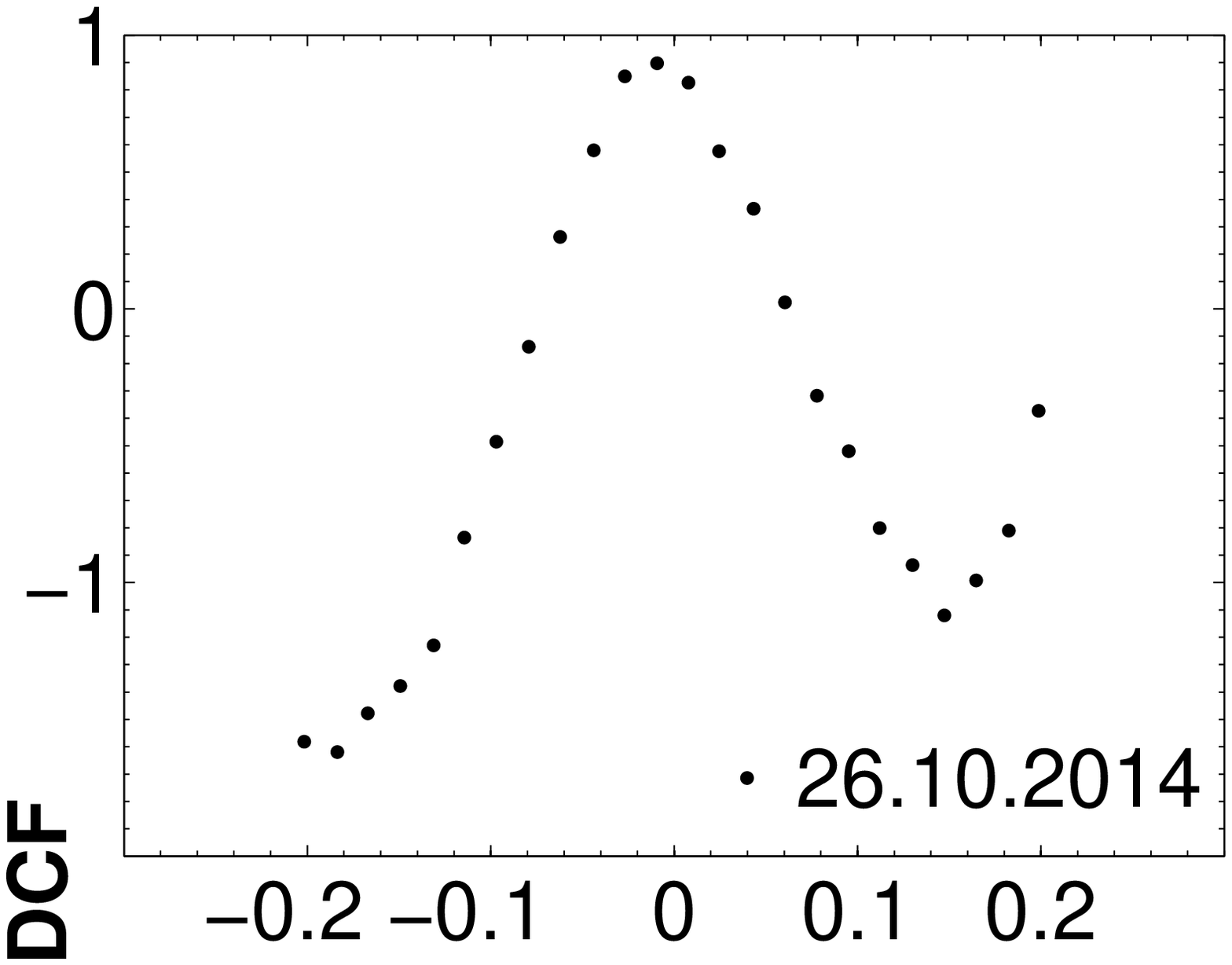,height=1.767in,width=1.865in,angle=0}
 \epsfig{figure=  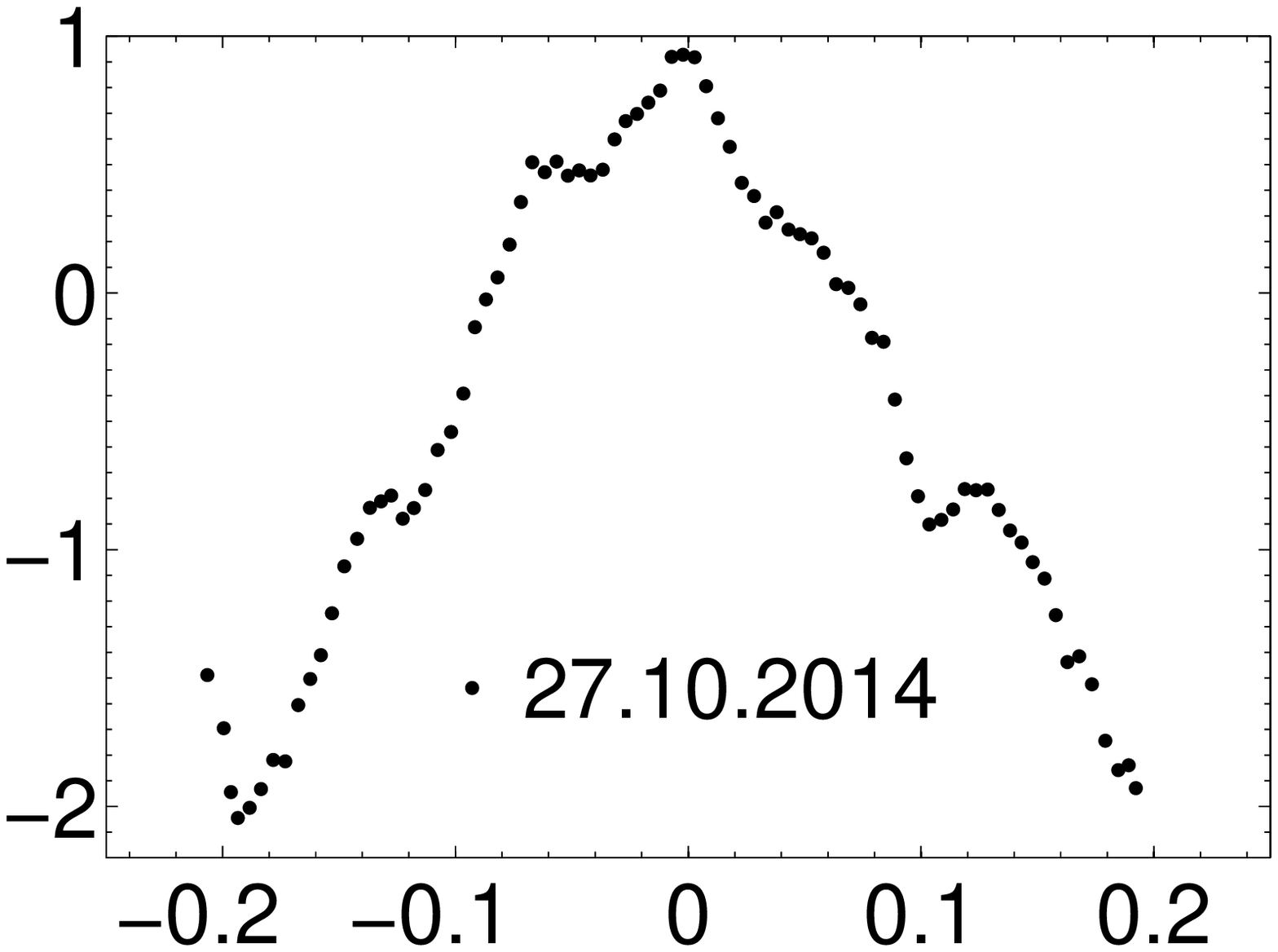,height=1.767in,width=1.865in,angle=0}
\epsfig{figure=  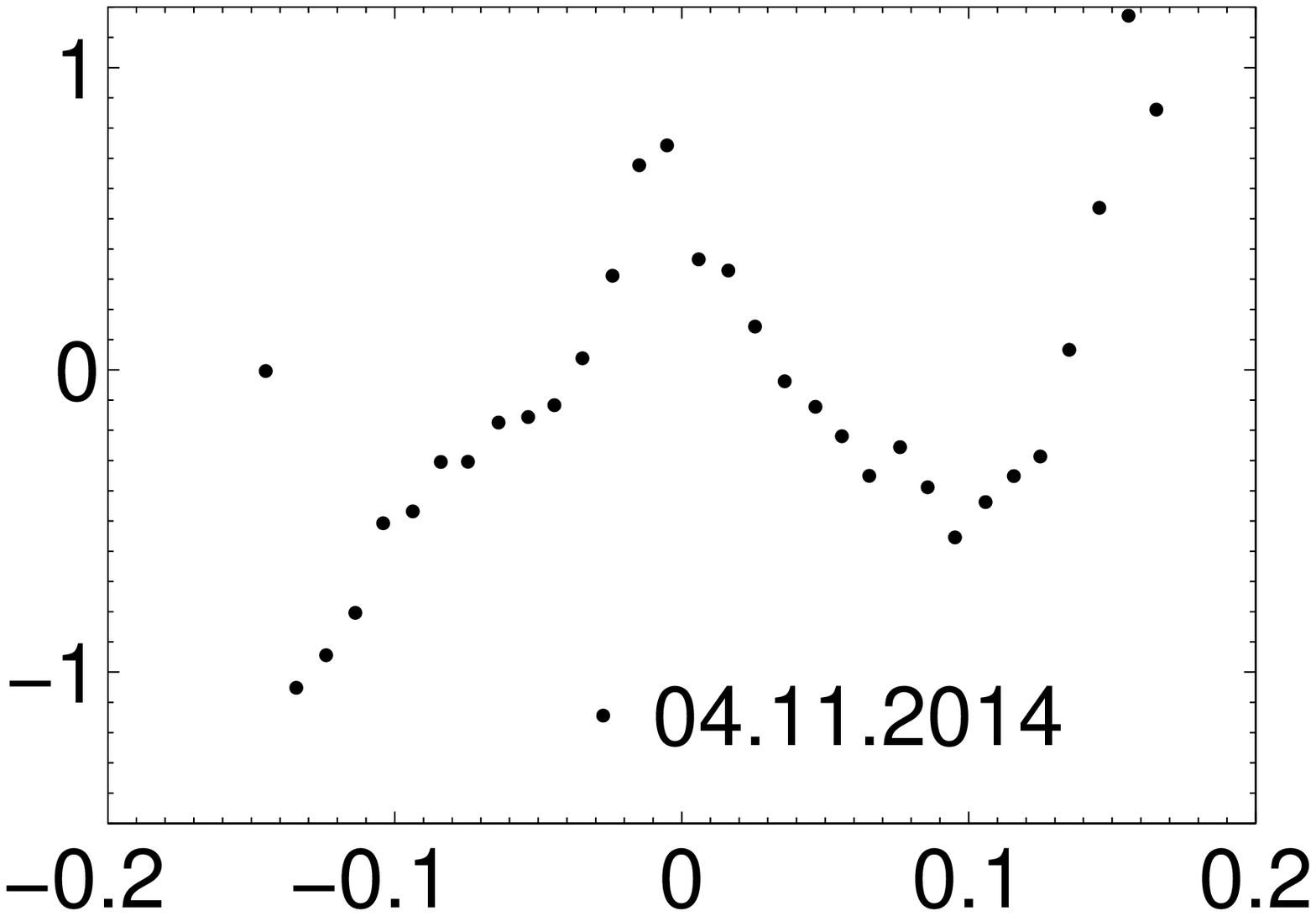,height=1.767in,width=1.865in,angle=0}
\epsfig{figure=  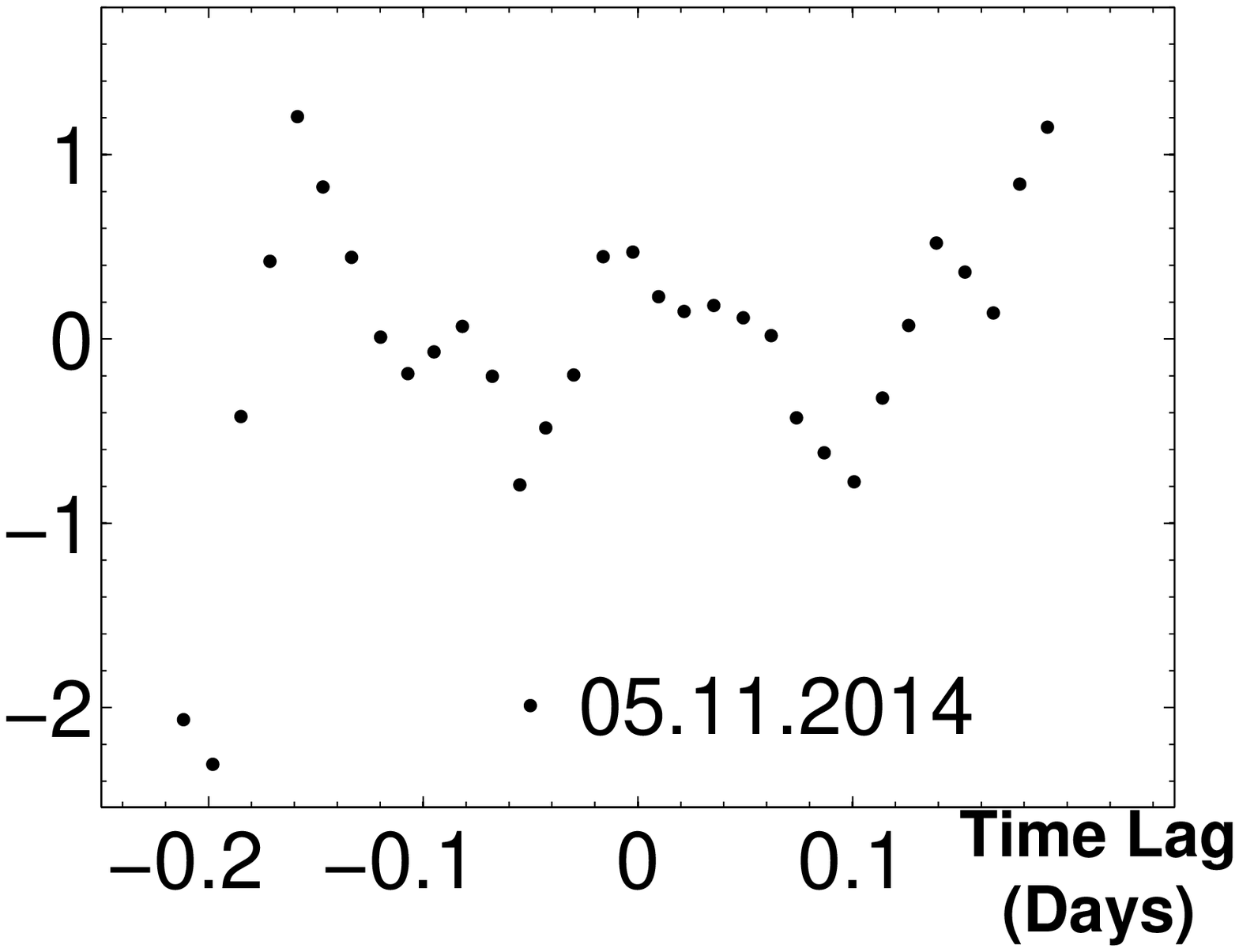,height=1.767in,width=1.865in,angle=0}
\epsfig{figure=  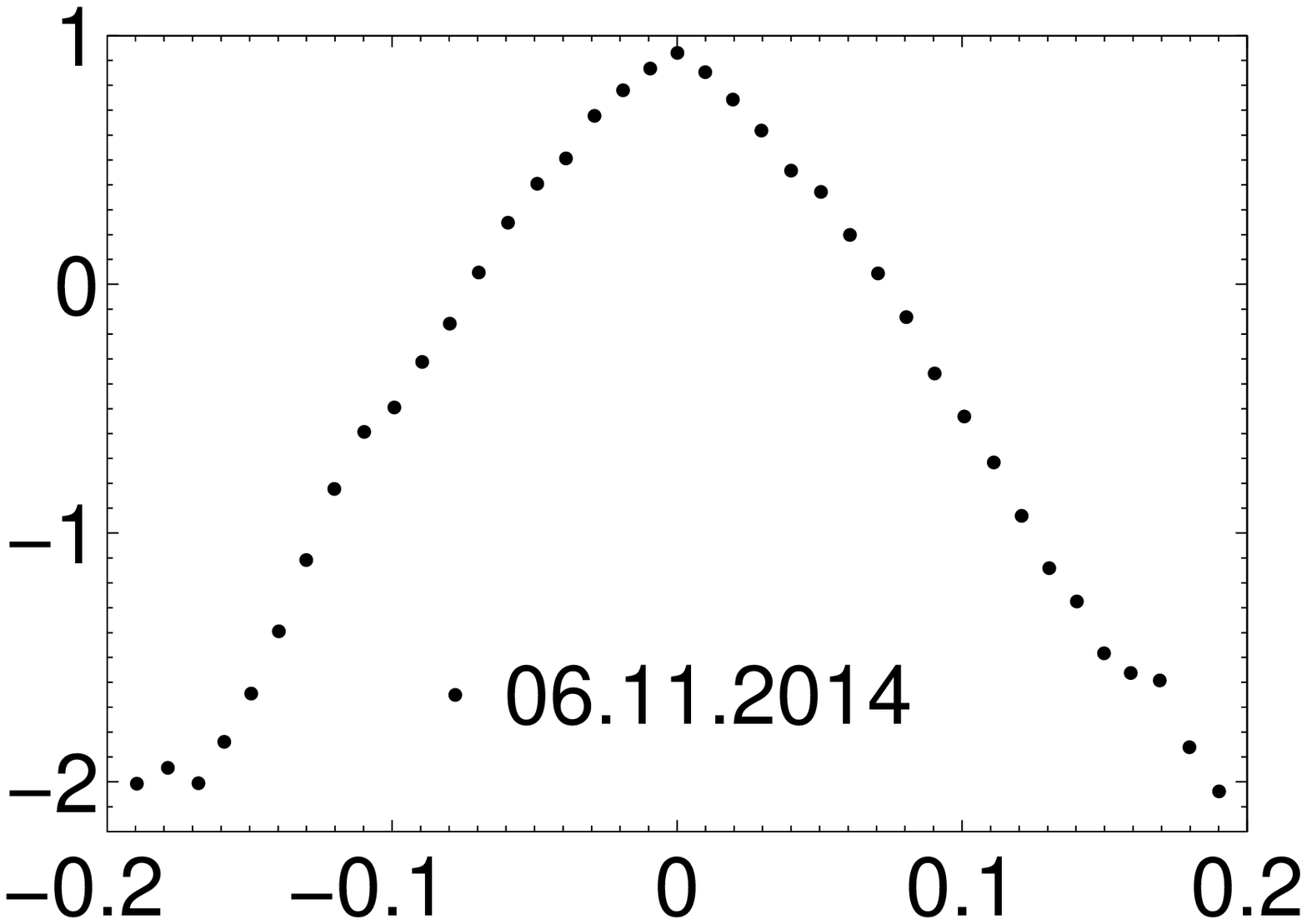,height=1.767in,width=1.865in,angle=0}
  \caption{ DCF plots between V and R passbands on intraday timescales for those nights when variability was detected in both the bands.
  The DCF values are given on
the Y-axis and the Time Lag is plotted against them for each labeled date of observation.}
\label{LC_BL}
\end{figure*} 

We used MATLAB software to write additional programs for data processing
We observed BL Lacertae on continuous 6 nights from 4 Nov to 9 Nov 2014 quasi-simultaneously in V and
R bands to study the IDV characteristics of the object and also to search for any possible time lags
between these bands. We took single data points in B and I bands also and found that the source
showed substantial brightness changes in all optical bands.

B, V, R and I light curves (LCs) of the BL Lacertae during our full monitoring period are displayed in Figure 1 where we fitted
smoothing splines (Bachev et al.\ 2011) on the light curve (LC) during continuous monitoring period between 4 Nov to 9 Nov, which
helps us to reveal the variability characteristics and time lags more clearly.
The upper panel of the Figure 1 represent (B-I) and (V-R) variations with time.

The observation log is given in Table 2 where we have listed observation date, telescope used and number of data points for each date in a particular
filter.
The source shows clear evidence of IDV on 3 nights in both V band and R band.

\section{Analysis Techniques}
To investigate variability properties of BL Lacertae we used a series of statistical analysis techniques to state that the extracted results are
statistically significant. All these statistical tools are developed using MATLAB. 

\subsection{\bf Variability detection criterion}

To quantify the intraday variability nature of the source we have employed three different statistics (e.g.,  de Diego 2010) named as C test, F test
and $\chi^{2}$test.

\noindent
\subsubsection{\bf C-Test}

To claim the variability of the source we used the most frequently used criterion introduced by Romero, Cellone, \& Combi (1999), where the variability
detection parameter C$_{avg}$ is defined as the average of $C_{1}$ and $C_{2}$ with:
\begin{equation}
C_{1} = \frac {\sigma(BL-Star A)}{\sigma(Star A-Star B )}~
,~
C_{2} = \frac {\sigma(BL-Star B)}{\sigma(Star A-Star B)}.
\end{equation}
Here (BL$-$Star A), (BL$-$Star B), and (Star A$-$Star B) are the differential instrumental magnitudes of 
the blazar and standard star A, the blazar and standard star B, and standard star A vs.\ standard star B 
determined using aperture photometry of the source and comparison stars, whereas $\sigma$(BL$-$starA), 
$\sigma$(BL$-$starB) and $\sigma$(Star A$-$Star B) are observational scatters of the differential instrumental 
magnitudes of the blazar$-$Star A, blazar$-$Star B, and Star A$-$Star B, respectively, with star A and star B being those two stars having
$\sigma(Star A-Star B)$ to be minimum. According to the adopted variability criterion, if $C_{avg} \geq 2.57$, then the nominal confidence level of variability
detection is $> 99$\% (Jang \& Miller 1997; Stalin et al.\ 2004; Gupta et al.\ 2008). If observations are done in three or more than three
filters then C$_{avg}$ value should be $\geq$ 2.576 in at least two filters for the source to be reported variable.

\subsubsection{\bf F-Test}
de Diego (2010) mentioned the C-test to be too conservative in quantifying variability. F test is considered to be a proper statistics to test
any changes of variability. F values compare two sample variances and are calculated as:
\begin{equation}
 \label{eq.ftest}
 F_1=\frac{Var(BL-Star A)}{Var(BL-Star B)}, \nonumber \\ 
 F_2=\frac{Var(BL-Star A)}{ Var(Star A-Star B)}.
\end{equation}
Here (BL-Star A), (BL-Star B), and (Star A-Star B) are the differential instrumental magnitudes of blazar and star A, blazar 
and star B, and star A and star B, respectively, while Var(BL-Star A), Var(BL-Star B), and Var(Star A-Star B) are 
the variances of differential instrumental magnitudes.

We take the average of $F_1$ and $F_2$ to find a mean observational F value.
The F value is then compared with $F^{(\alpha)}_{\nu_{bl},\nu_*}$, a critical value, where $\nu_{bl}$ and $\nu_*$ 
respectively denote the number of degrees of freedom for the blazar and star, while $\alpha$ is the significance level set as
0.1 and 1 percent (i.e $3 \sigma$ and $2.6 \sigma$) for our analysis. If the mean F value is 
larger than the critical value, the null hypothesis (i.e., that of no variability) is discarded.
Nightly LCs and colour indices are listed as variable if $F > F_c(0.99)$.

\begin{figure*}
\epsfig{figure=  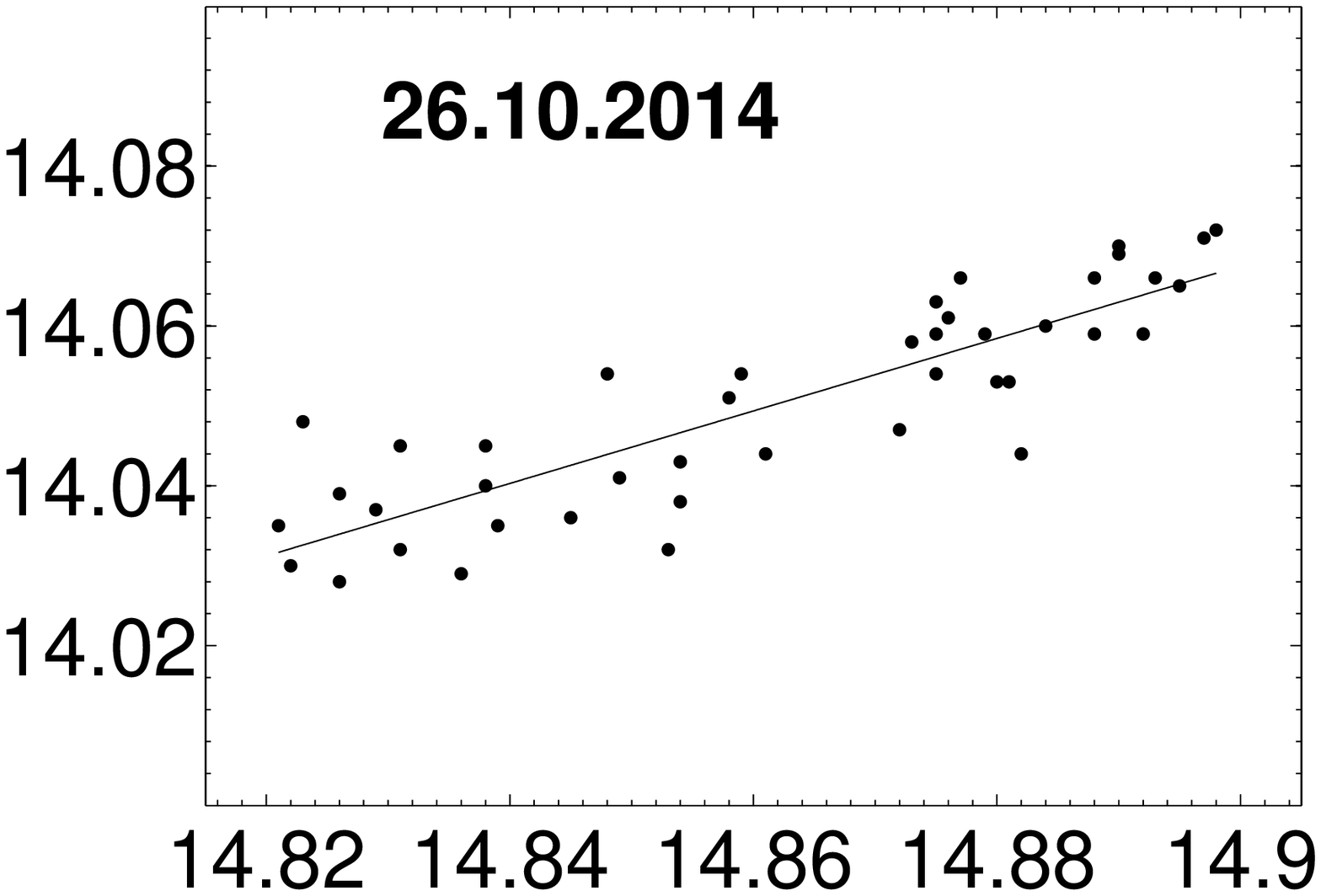,height=1.267in,width=1.365in,angle=0}
 \epsfig{figure=  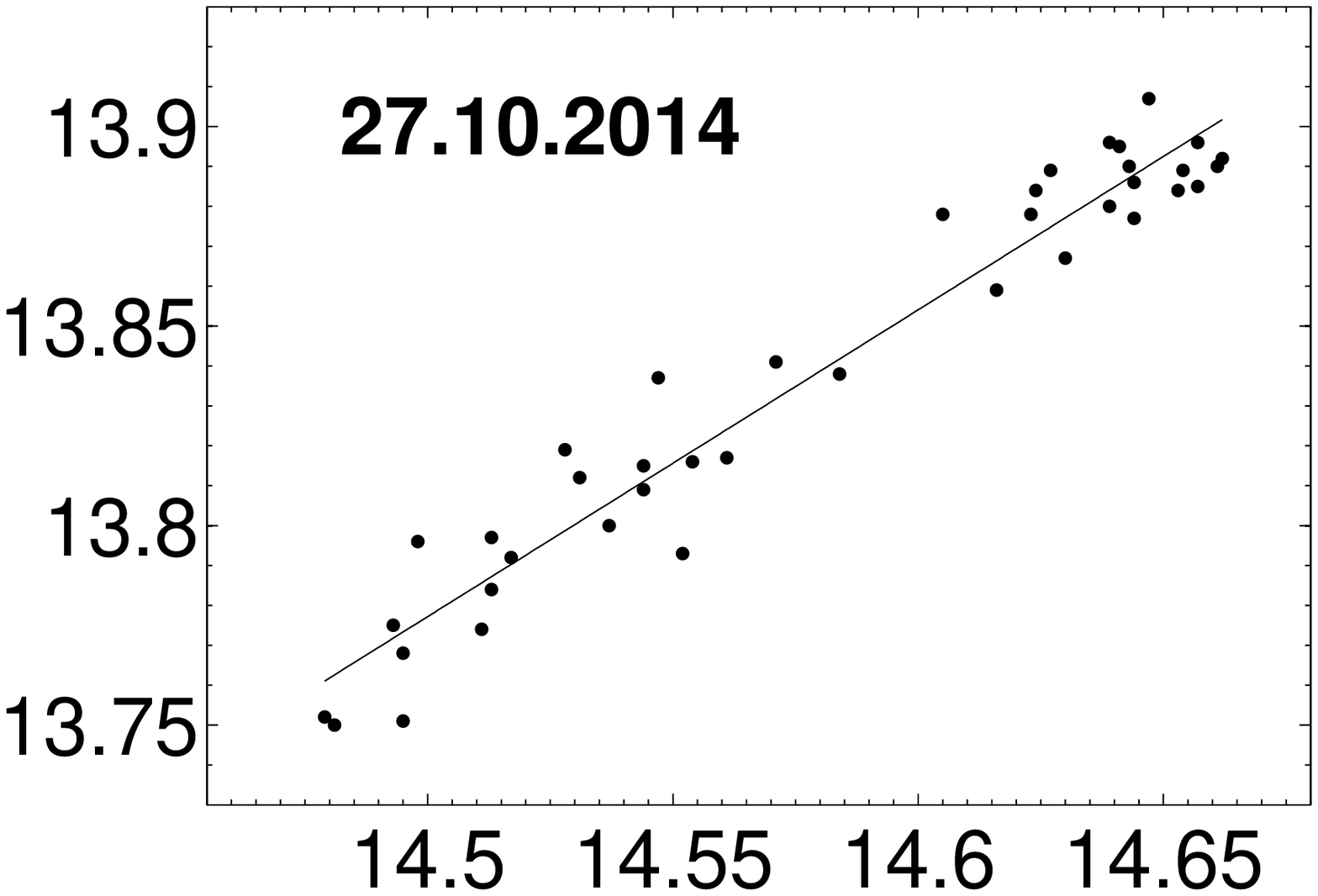,height=1.267in,width=1.365in,angle=0}
\epsfig{figure= 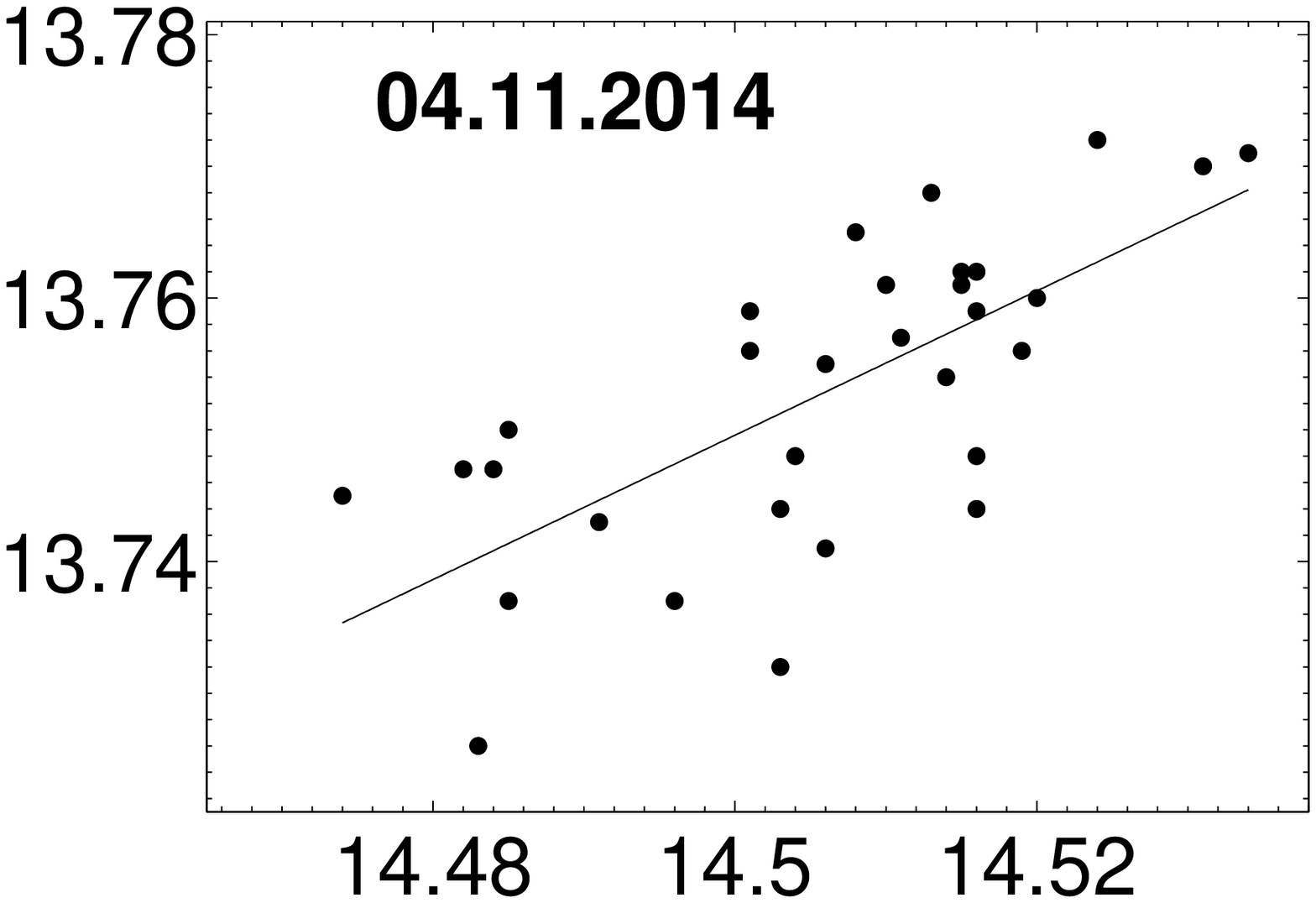 ,height=1.267in,width=1.365in,angle=0}
\epsfig{figure=  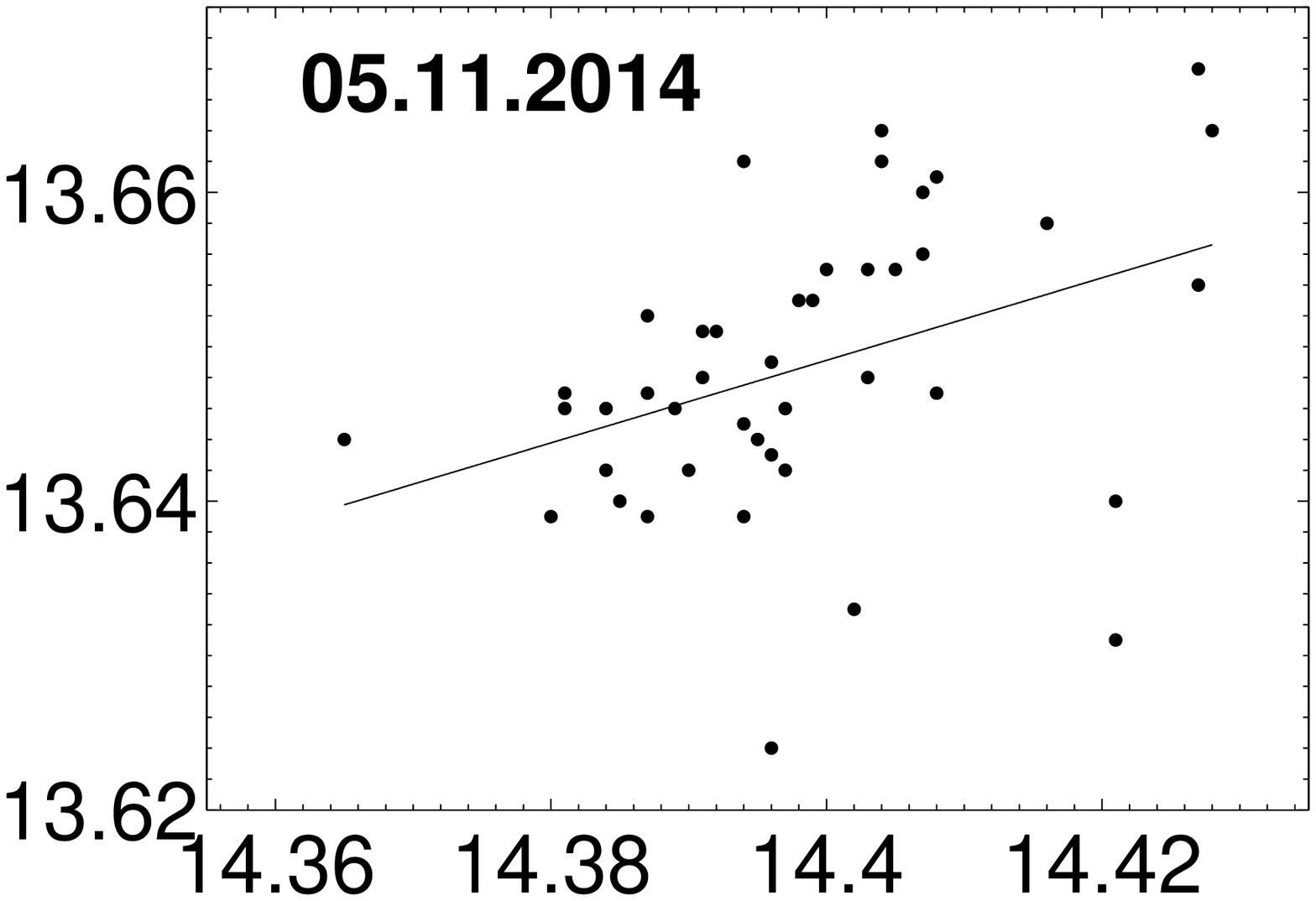,height=1.267in,width=1.365in,angle=0}
\epsfig{figure=  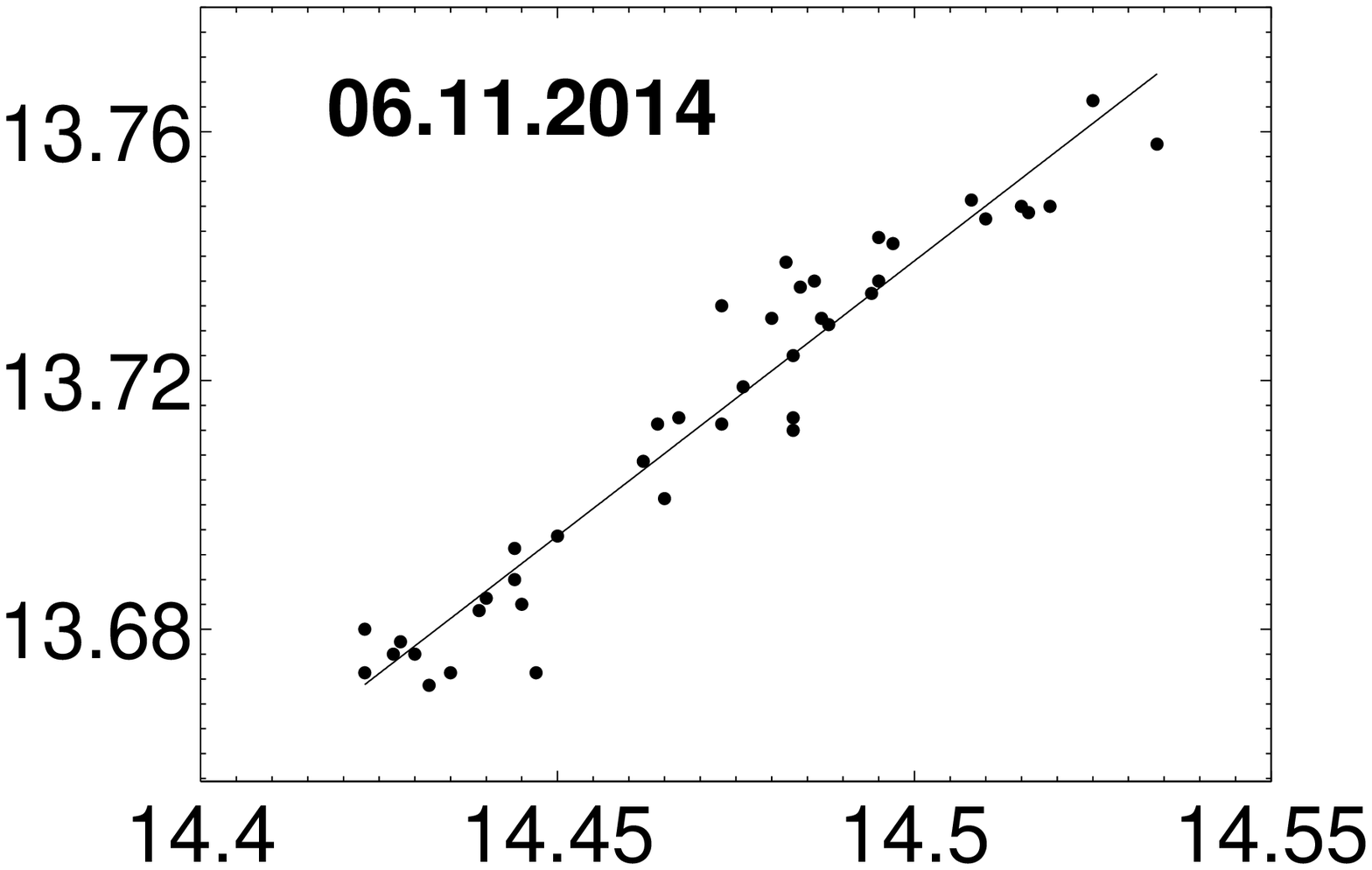,height=1.267in,width=1.365in,angle=0}
\epsfig{figure=  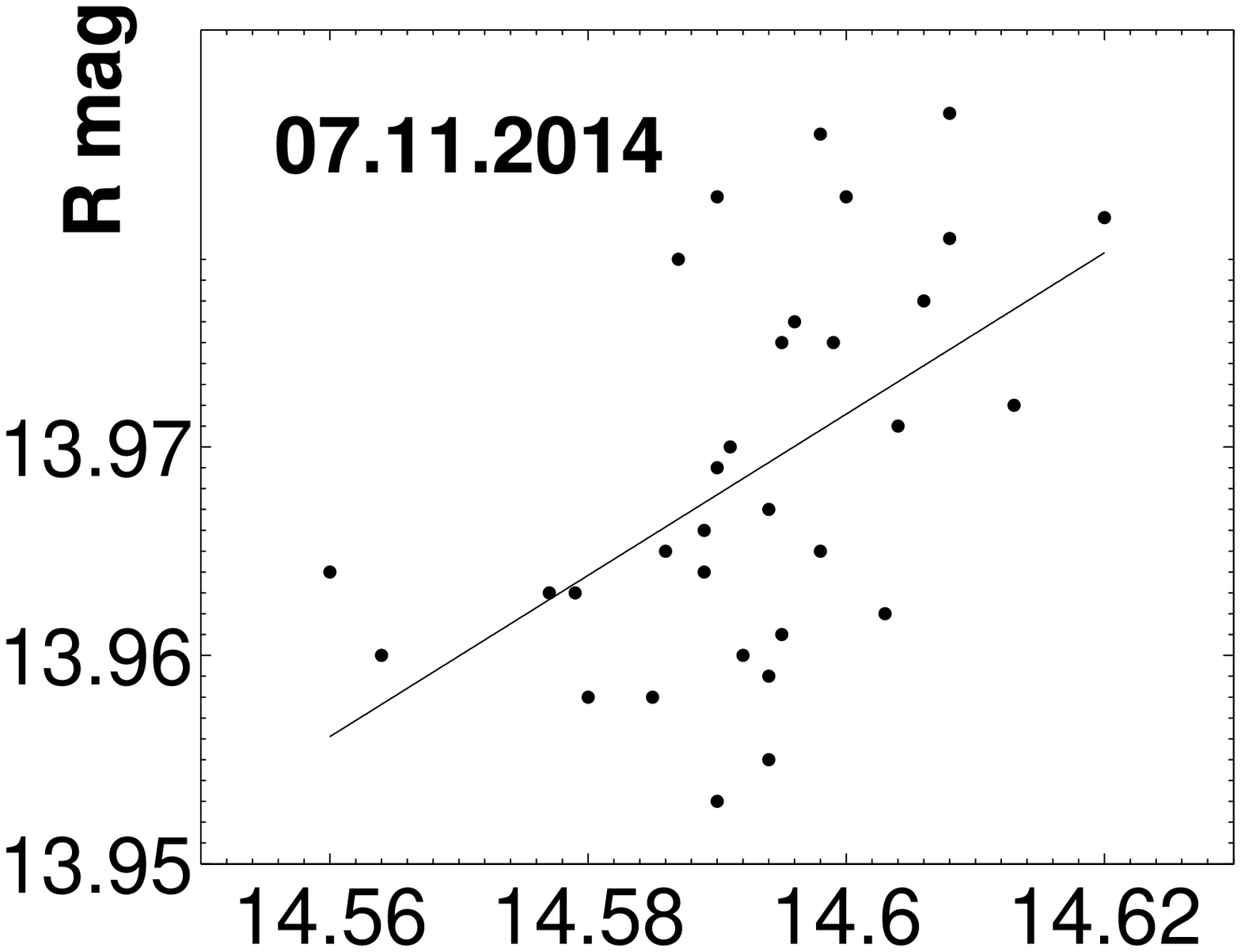,height=1.267in,width=1.365in,angle=0}
\epsfig{figure=  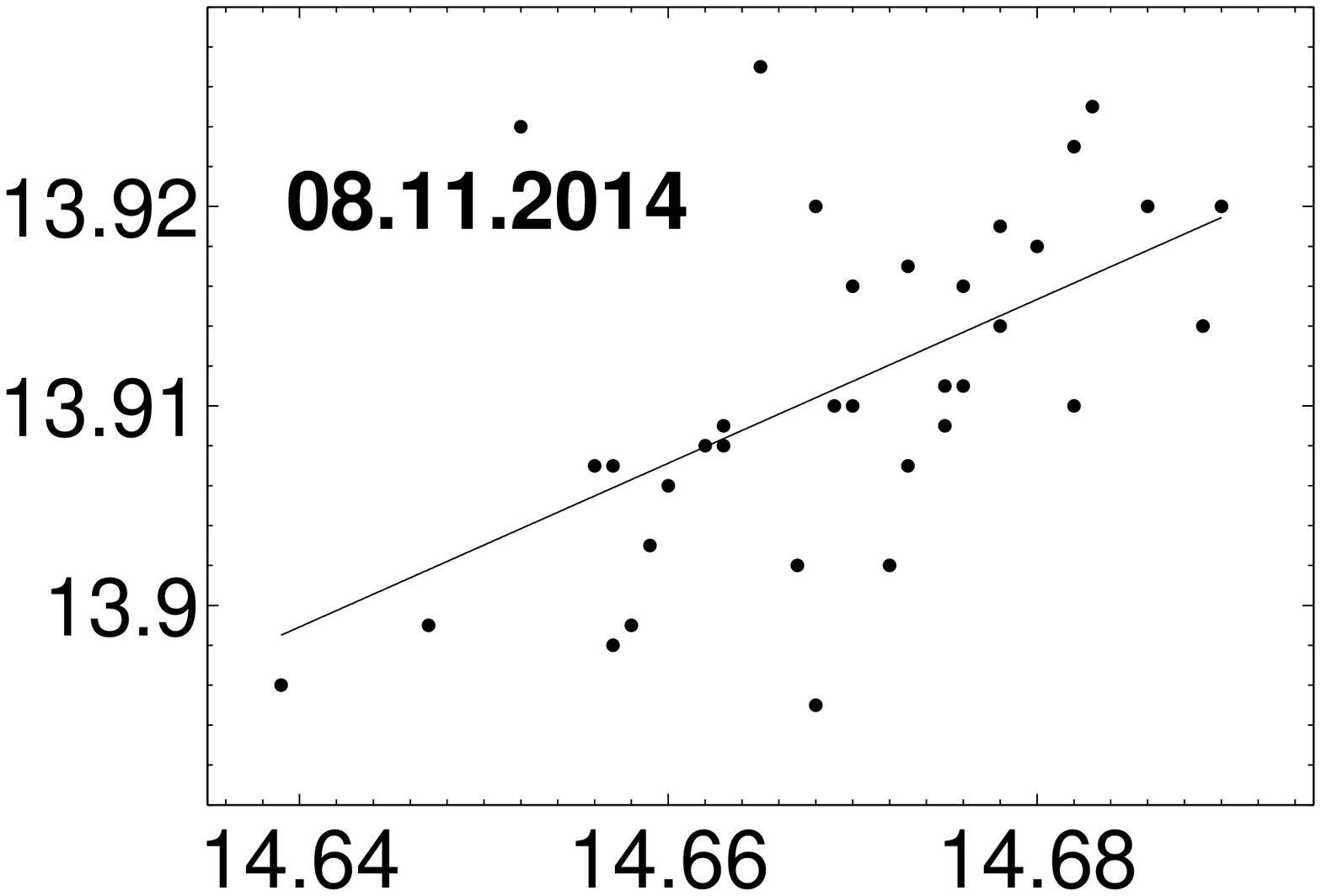,height=1.267in,width=1.365in,angle=0}
\epsfig{figure= 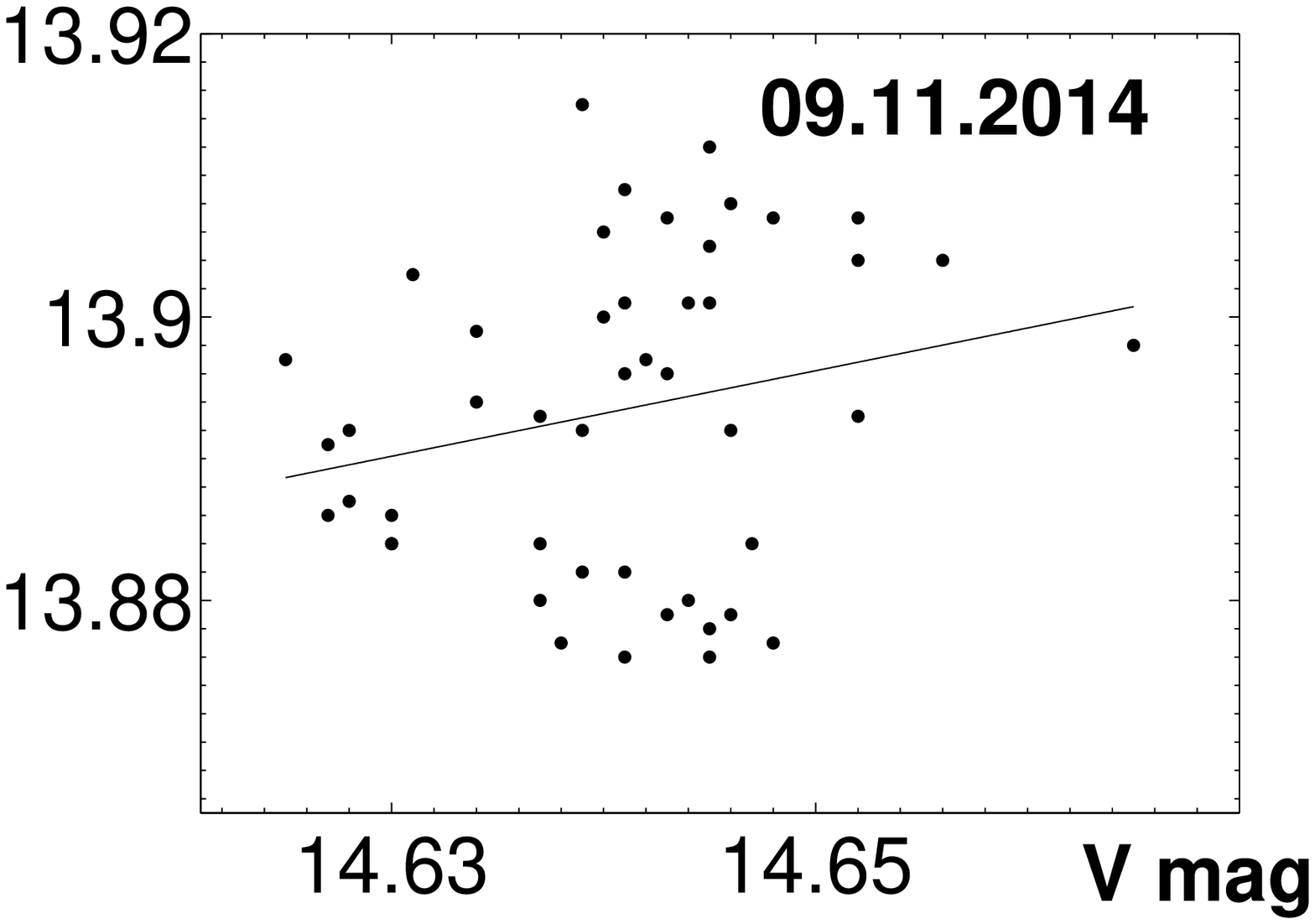,width=1.365in,height=1.267in,angle=0}
 \epsfig{figure= 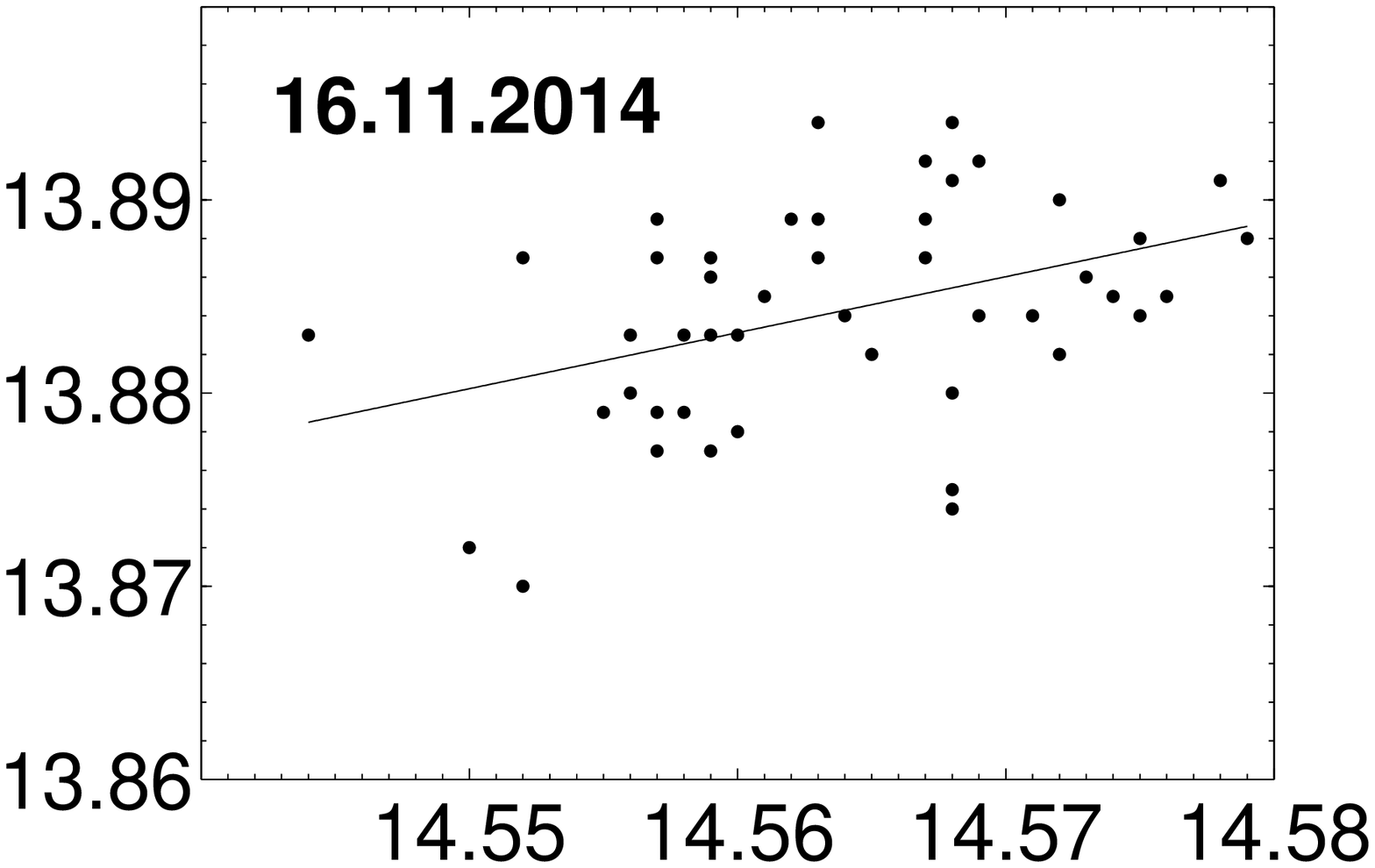,width=1.365in,height=1.267in,angle=0}
  \epsfig{figure= 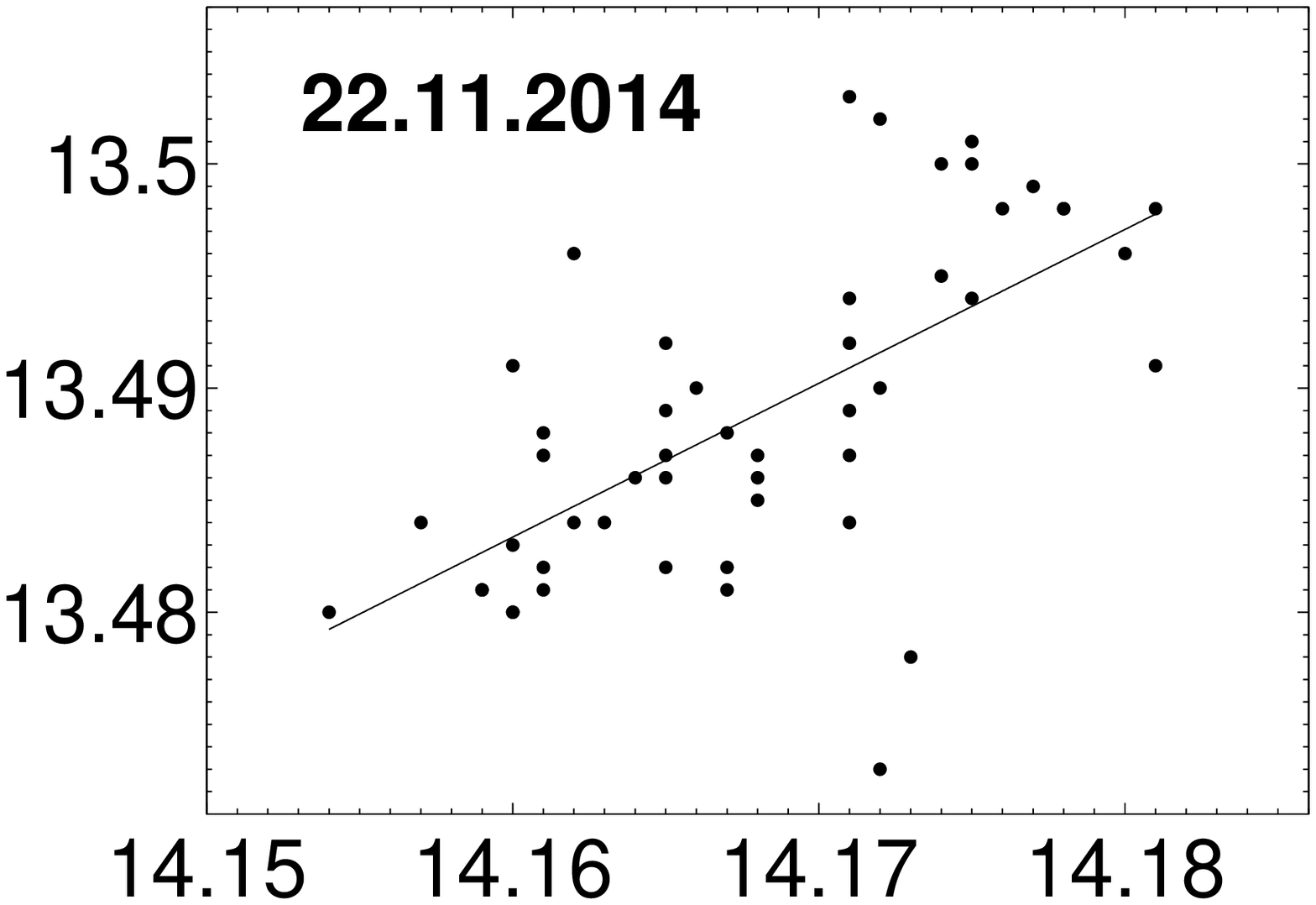,width=1.365in,height=1.267in,angle=0}
  \caption{R magnitude (y-axis) Vs V magnitude (x-axis) plots on intra-night timescale for 10 observation days.}
\label{LC_BL}
\end{figure*}

\subsubsection{\bf $\chi^{2}$-test}

To investigate the presence of IDV we also performed a $\chi^{2}$-test. The $\chi^{2}$ statistic is defined as

\begin{equation}
\chi^2 = \sum_{i=1}^N \frac{(V_i - \overline{V})^2}{\sigma_i^2},
\end{equation}
where, $\overline{V}$ is the mean magnitude, and the $i$th observation yields a magnitude $V_i$
with a corresponding standard error $\sigma_i$ which is due to photon noise from the source and sky, CCD read-out and other non-systematic
error sources. 
Exact quantification of such errors by the IRAF reduction package is impractical and it has been found that
theoretical errors are smaller than the real
errors by a factor of 1.3-1.75 (e.g., Gopal-Krishna et al.\ 2003, Gupta et al.\ 2008) which for our data is $\sim$1.5, on average. 
So the errors obtained after data reduction 
should be multiplied by this factor to get better estimates of the real photometric errors.
This statistic is then compared with a critical value $\chi_{\alpha,\nu}^2$ where $\alpha$ is again the significance level set same as in case of
F test while $\nu = N -1$ is the degree of freedom. $\chi^2 > \chi_{\alpha,\nu}^2$ implies the presence of variability.

\noindent
\subsubsection{\bf Percentage amplitude variation}

The percentage variation on a given night is calculated by using the variability amplitude parameter $A$,
introduced by Heidt \& Wagner (1996), and defined as
\begin{eqnarray}
A = 100\times \sqrt{{(A_{max}-A_{min}})^2 - 2\sigma^2}(\%) ,
\end{eqnarray}
where $A_{max}$ and $A_{min}$ are the maximum and minimum values in the calibrated LCs of the blazar, and $\sigma$
is the average measurement error.

\subsection{\bf Discrete correlation function (DCF)}
To search for and quantify any possible correlations between the optical fluxes we used the DCF technique.
The DCF method was first introduced by Edelson \& Krolik (1988). For details on DCF see Hovatta et al.\ (2007); Rani, Wiita, \& Gupta (2009), and
references therein.

The DCF method is defined as: for each pair of data \( (x_i , y_j ) \), with 
$0 \leq i,j \leq N$, with $N$  the number of data points, the unbinned discrete correlation function 
(UDCF) is
\begin{eqnarray}
 UDCF_{ij}(\tau) = \frac{(x_i-\bar{x})(y_j-\bar{y})}{\sqrt{(\sigma_{x^2} - e_{x^2})(\sigma_{y^2} - e_{y^2})}}
\end{eqnarray}
where the parameters $\bar{x}, \bar{y}$ are the mean values of the two discrete data series \( x_i , y_j \),
with standard deviations \( \sigma_x\), \( \sigma_y \) and measurement errors $e_x$, $e_y$.

The DCF can be calculated by averaging the UDCF values ($M$ in number) for each time delay 
\( \Delta t_{ij} = (t_{yj}-t_{xi}) \) lying in the range  \( \tau - \frac{\Delta\tau}{2} \leq t_{ij} 
\leq \tau+ \frac{\Delta\tau}{2} \) and is expressed as
\begin{equation}
 DCF{(\tau)}= \frac{\sum_{k=1}^m UDCF_{k}}{M} ,
\end{equation}
where $\tau$ is the center of the bin of size $\Delta \tau$. The error is found from the standard deviations
of the number of bins used for determining the DCF and is given as:
\begin{equation}
 \sigma_{DCF(\tau)} = \frac{\sqrt{\sum_{k=1}^{M} (UDCF_k-DCF(\tau))^2}}{M-1} .
\end{equation}

When correlating a data series with itself $(x=y)$, we obtain the auto-correlation function (ACF) with a peak
at $\tau =0$, indicating the absence of any time lag while any other strong peak indicates the presence of periodicity.
In general, a DCF value $> 0$ 
implies the two data signals are correlated, while the two anti-correlated data sets have a DCF $<$ 0, and a DCF 
value equal to 0 implies no correlation exists between the two data trains. Additional advantage of this method is that it is suitable for
unevenly sampled data as is the case in most astronomical observations.

\section{\bf Flux and colour variability Results}

We observed the blazar BL Lacertae on 13 nights in B, V, R, and I passbands which includes quasi-simultaneous observations in V and R passbands on
10 of these 13 nights. The complete observation log for this blazar is given in Table 2.
Intra-night LCs in V and R filters taken quasi-simultaneously are plotted in Figure 2.
We applied C-statistics, F-test and $\chi^{2}$-test, as discussed above. Our analysis results on intranight timescales for
V and R filters are summarized in Table 3.
The blazar is marked as variable (Var) if the variability conditions for each test mentioned under Section 3.1 are satisfied, while its marked
probably variable (PV) if conditions for either of the two tests are followed, and finally the quasar is marked Non-variable (NV) if none of
the conditions are met by the target.
The source was found to be in active state during our entire span of observations with clear variability found on 3 nights as evident from the
Table 3 on 26 Oct, 27 Oct and 6 Nov 2014 in both V and R bands.
There was a prominent variation on 27 October 2014 around JD 245658.35 with amplitude of about $\sim$ 0.05 mag
followed by another short flaring event around JD 245658.43 of $\sim$ 0.04 mag in R passband. Similar trend was found in V passband
as shown in Figure 2
from which it is evident that the data sets for these events in both the bands are consistent with each other.
The object reached the faintest level of 14.07 mag in R passband which is still $\sim$ 1 mag brighter than its historical maximum magnitude as
reported by Fan \& Lin (2000). While on the last day of our observation span i.e., on 22 November,
source was found to be the brightest with a magnitude of 15.08, 14.15, 13.4, and
1.624 in B, V, R, and I passbands, respectively, indicating that the source might go into flaring state in near future. 
The intra-night variability amplitude was minimum on 22 Nov with value 2.96\% in R filter while reached maximum value of 18.28\% on 27 Oct in
V passband when source showed $\sim$0.18 mag change in 3.5 hours of observation duration in V passband.
As can be seen from the Table 3 the amplitude of variability is larger in V band LCs than the R passband LCs. The amplitude of variations is
mostly found to increase with frequency thus implying that the source spectrum becomes flatter when brightness of source increases
(Massaro et al.\ 1998) which is also the case with our target.

\begin{figure}
\epsfig{figure= 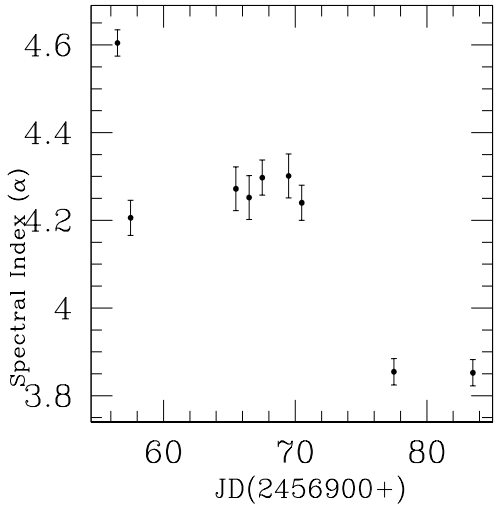,height=2.7in,width=2.6in,angle=0}
\caption{Variation of average optical spectral index calculated using equation 8 Vs time covering the entire observation period for the BL Lacertae.}
\label{LC_BL}
\end{figure}

The average spectral indices quoted in Table 4 are calculated simply as (Wierzcholska et al.\ 2015),
\begin{equation}
\langle \alpha_{VR} \rangle = {0.4\, \langle V-R \rangle \over \log(\nu_V / \nu_R)} \, ,
\end{equation}
where $\nu_V$ and $\nu_R$  are effective frequencies of the respective bands (Bessell, Castelli, 
\& Plez 1998).
The optical continuum of the source was very steep on most of the nights with $\alpha_{VR}$ $>$ 3.8
indicating strong synchrotron emission from the blazar jet and very less accretion disc contribution. Variation of spectral index with
time is shown in Figure 5 giving the average spectral index during the entire
observation cycle as $\sim$ 4.2.
\begin{figure*}
\epsfig{figure=  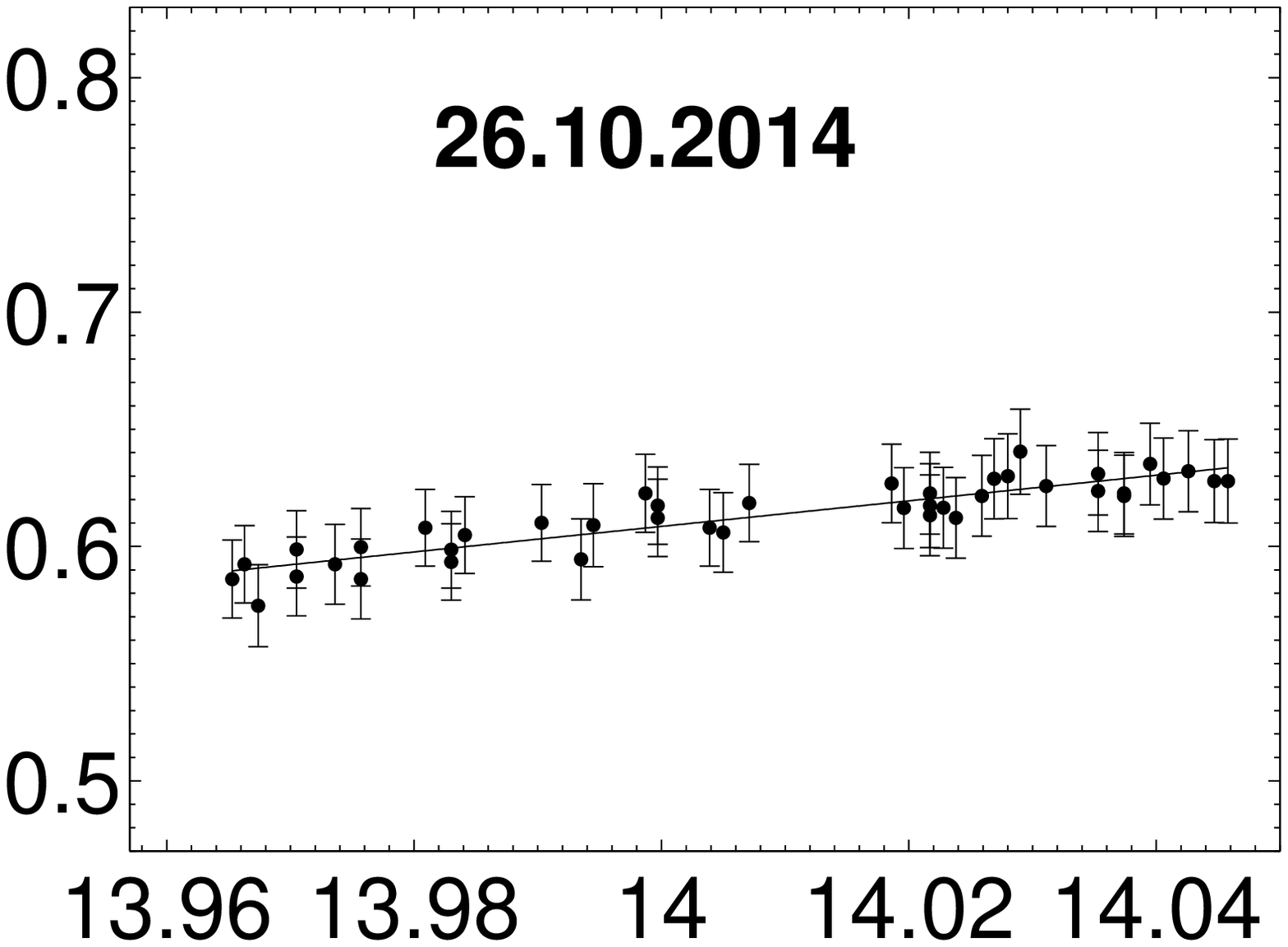,height=1.267in,width=1.365in,angle=0}
 \epsfig{figure=  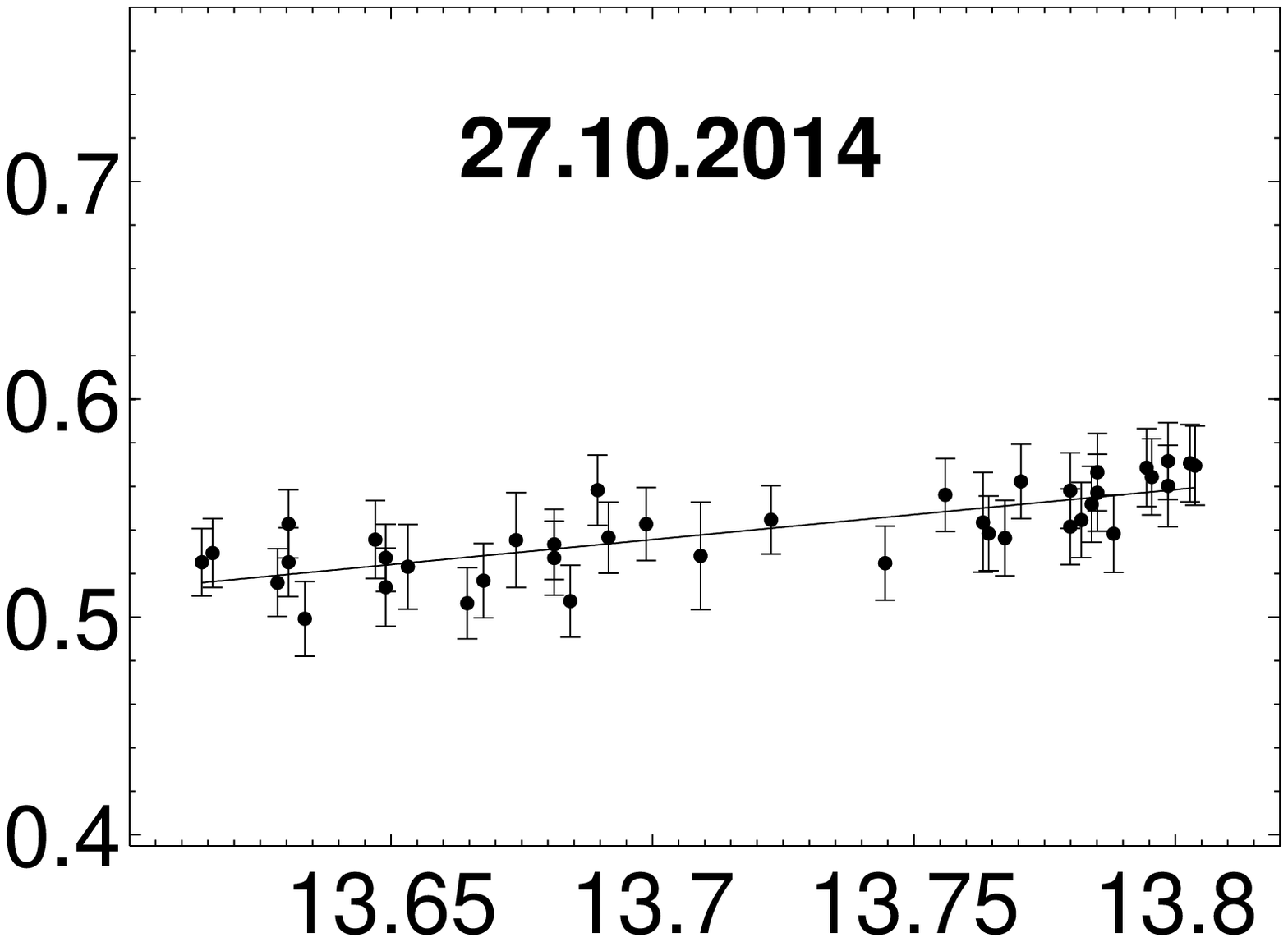,height=1.267in,width=1.365in,angle=0}
\epsfig{figure= 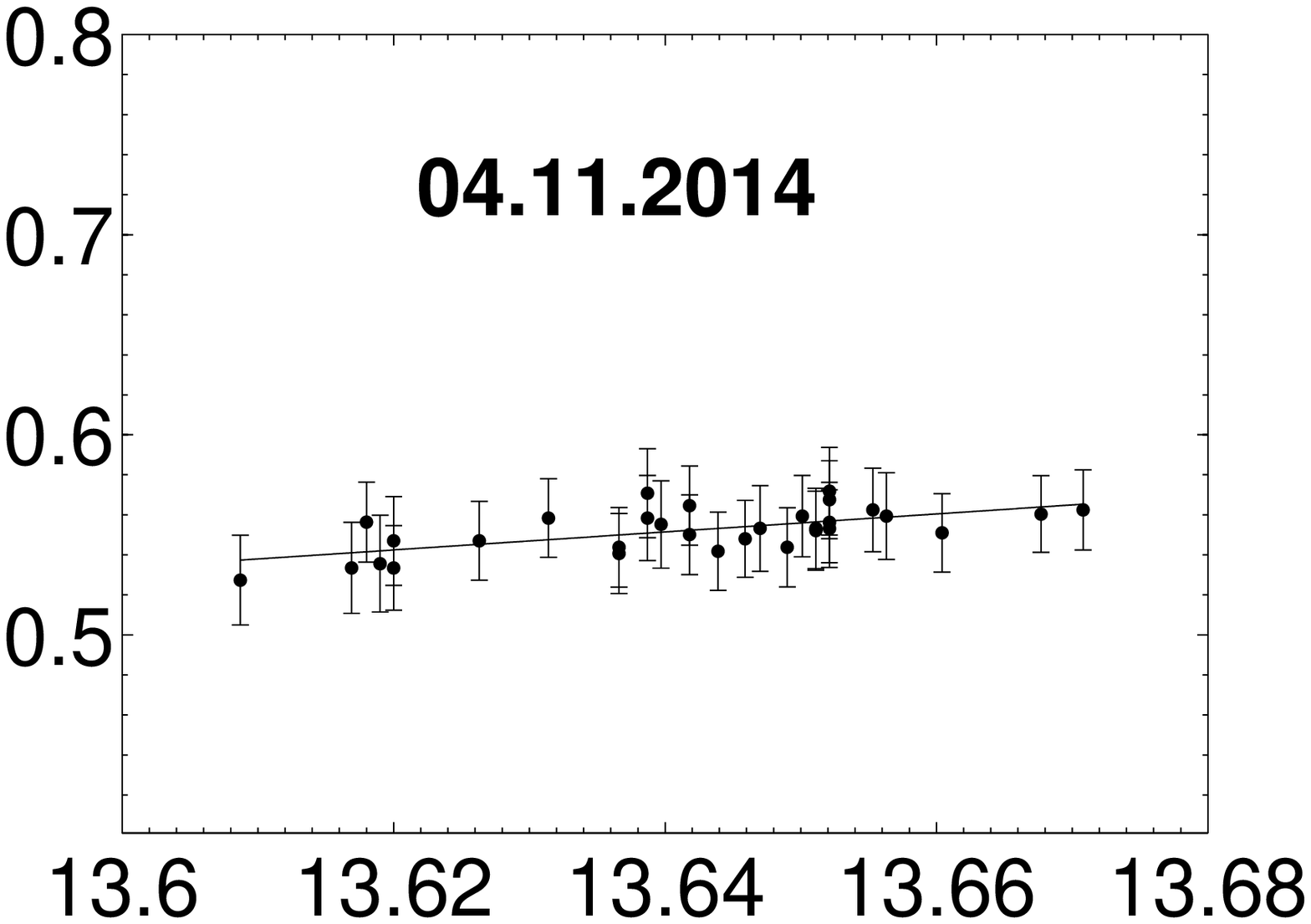 ,height=1.267in,width=1.365in,angle=0}
\epsfig{figure=  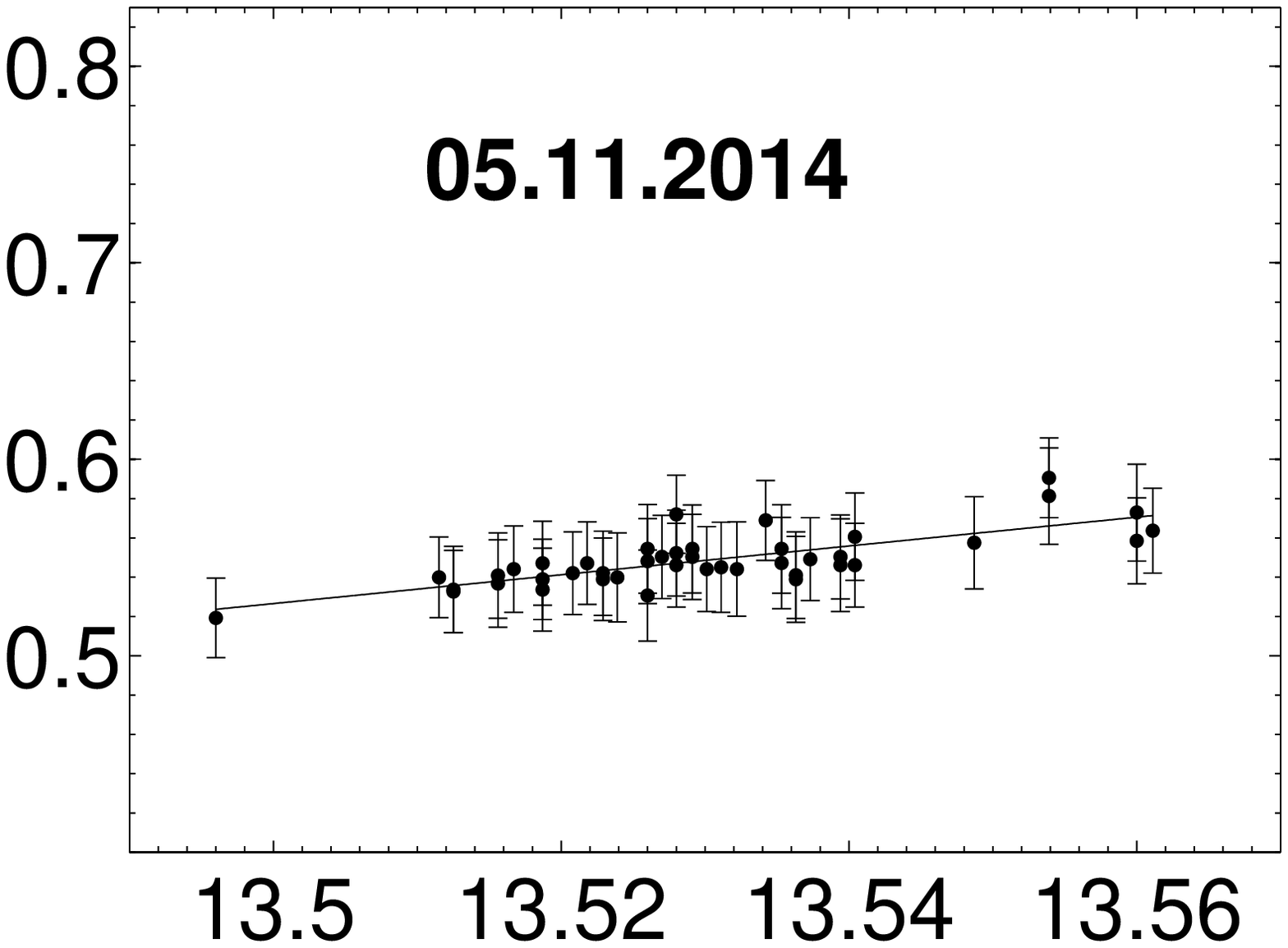,height=1.267in,width=1.365in,angle=0}
\epsfig{figure=  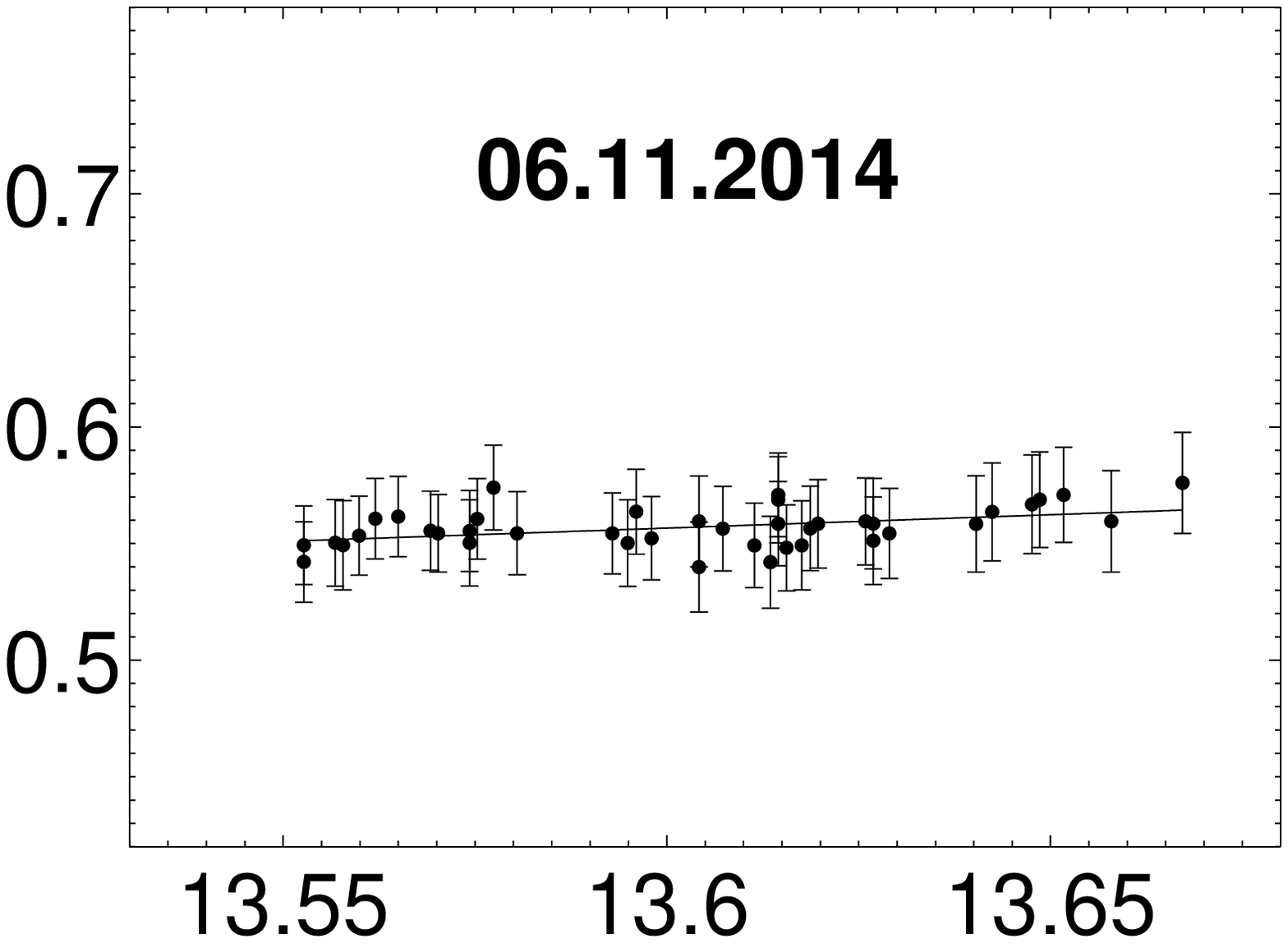,height=1.267in,width=1.365in,angle=0}
\epsfig{figure=  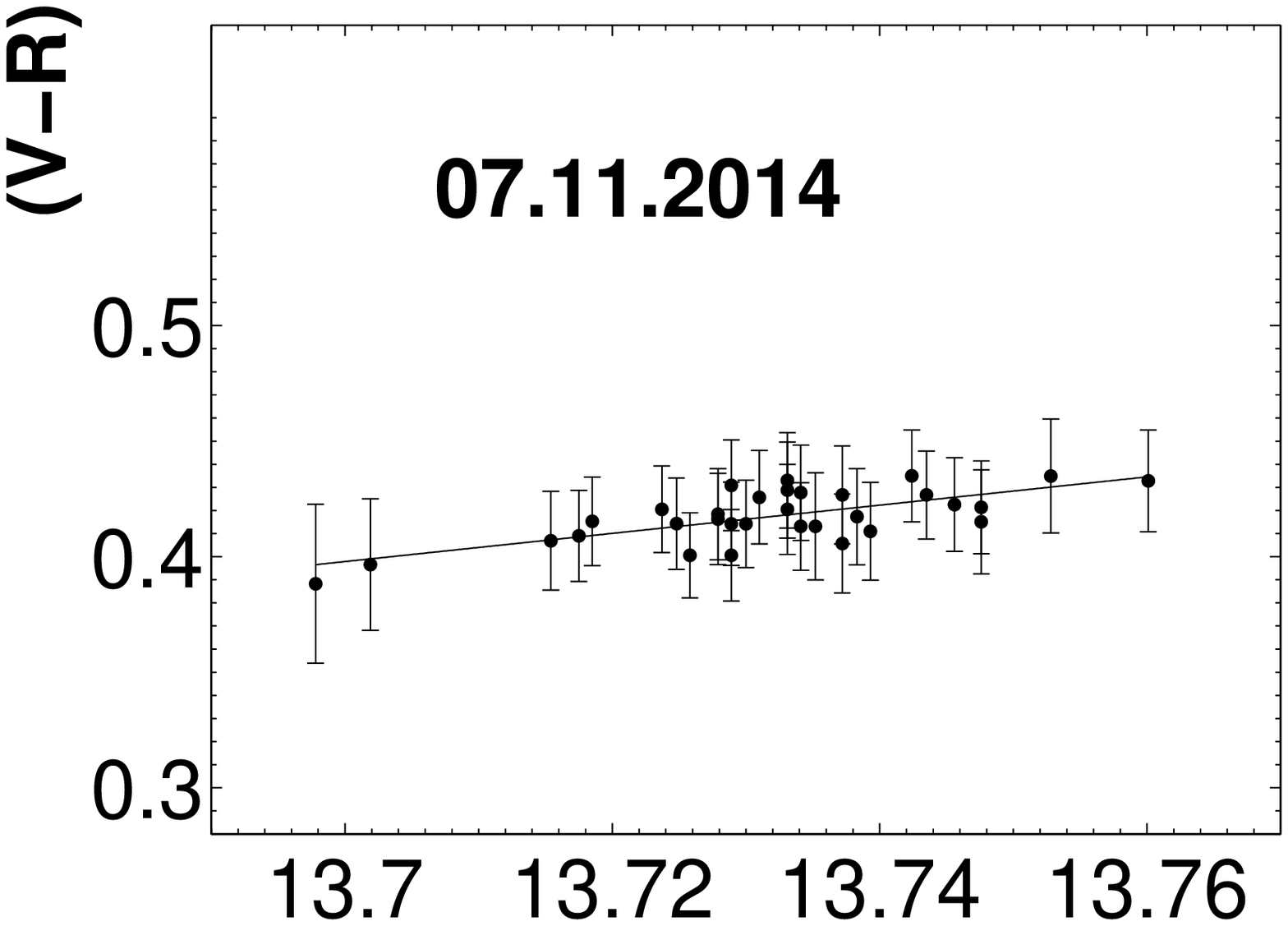,height=1.267in,width=1.365in,angle=0}
\epsfig{figure=  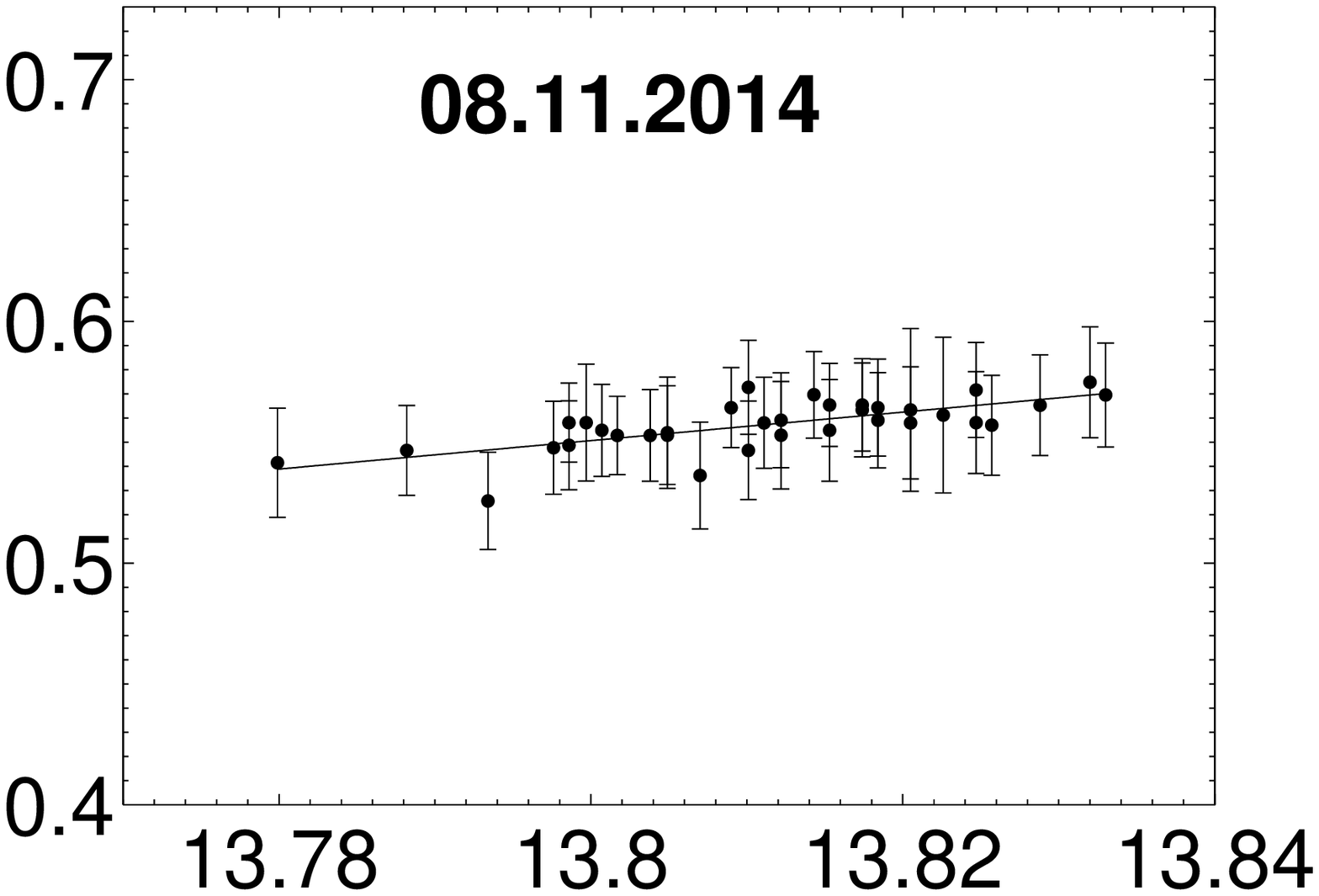,height=1.267in,width=1.365in,angle=0}
\epsfig{figure= 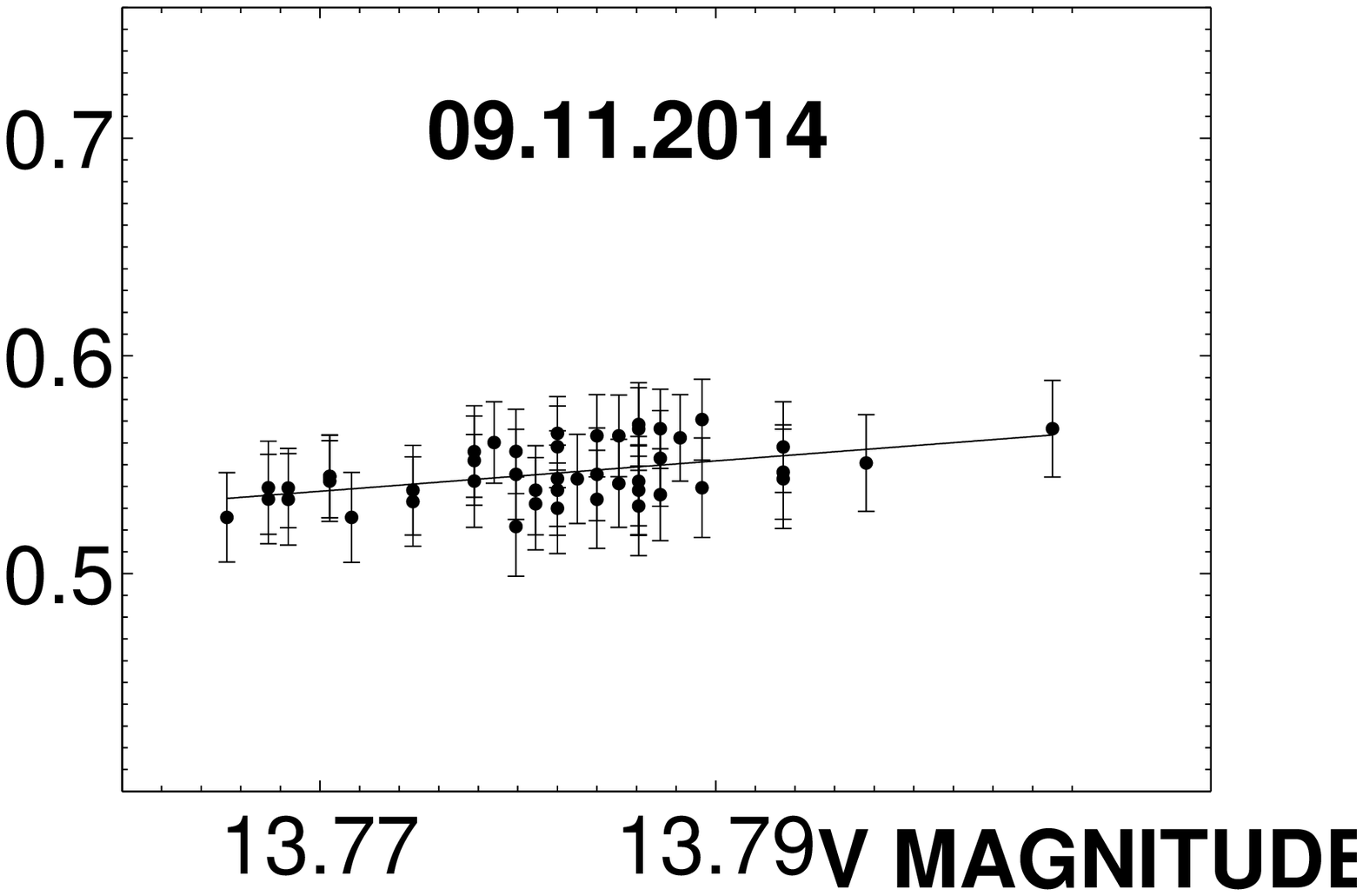,width=1.365in,height=1.267in,angle=0}
 \epsfig{figure= 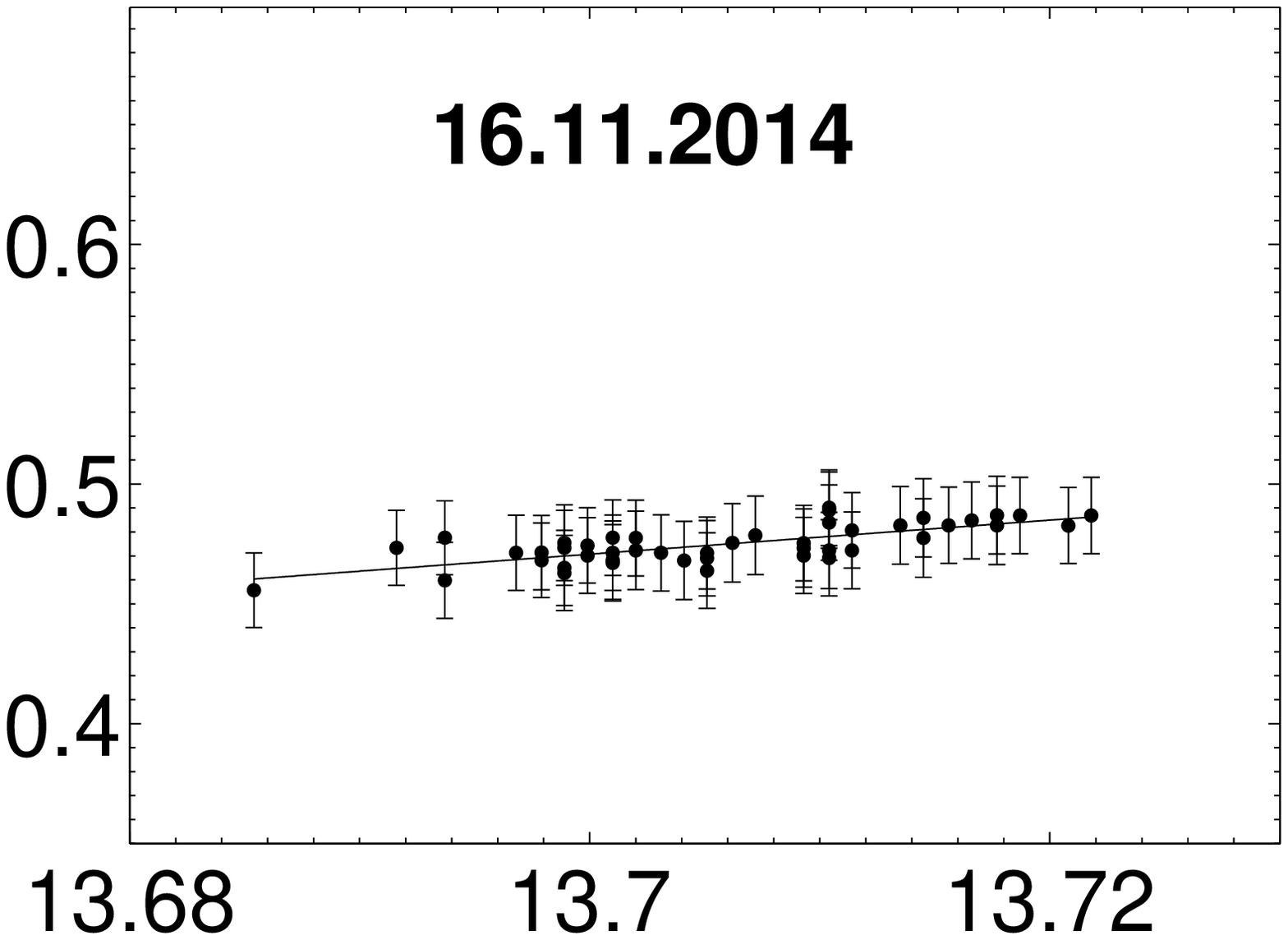,width=1.365in,height=1.267in,angle=0}
  \epsfig{figure= 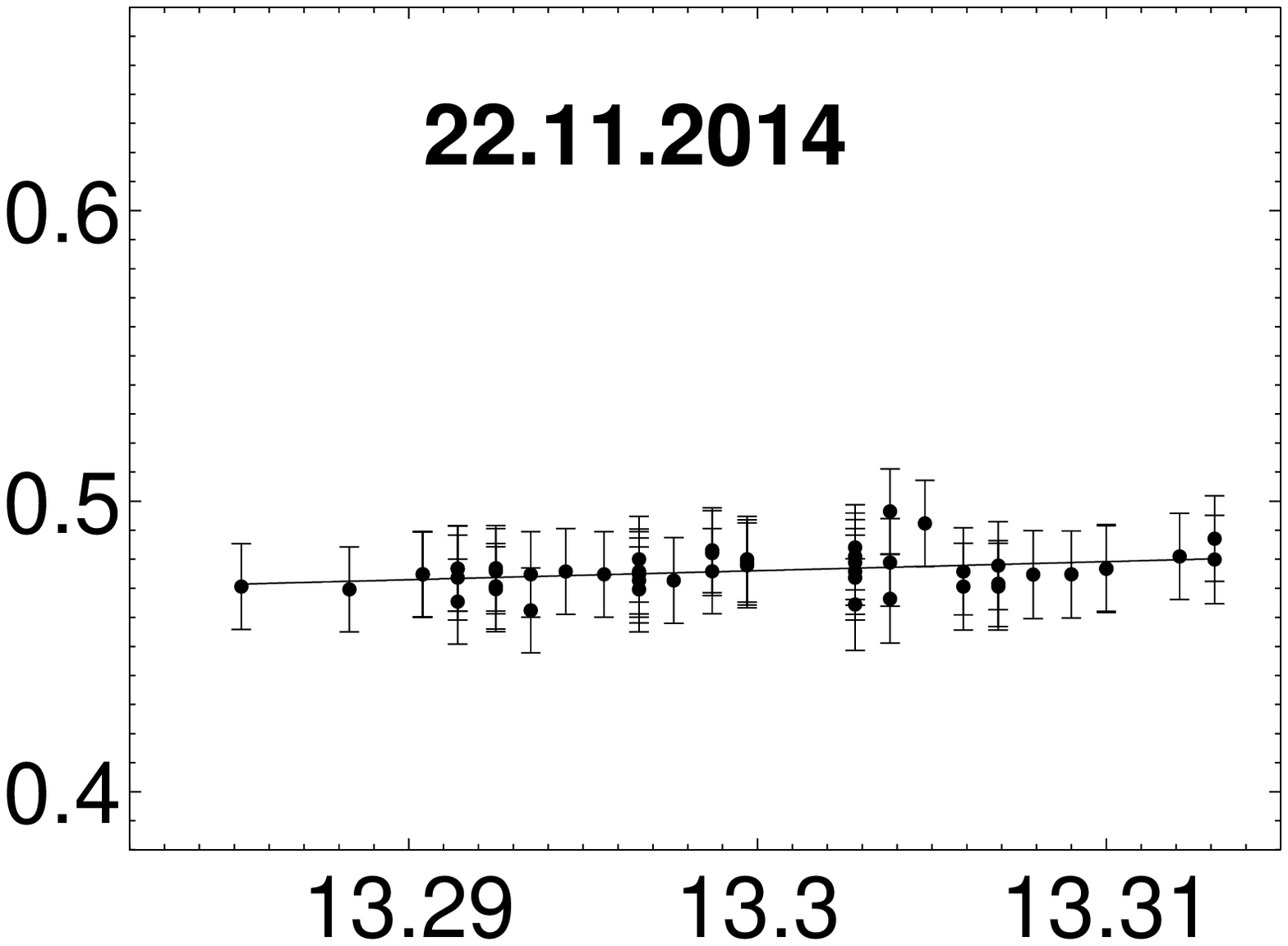,width=1.365in,height=1.267in,angle=0}
  \caption{ Colour magnitude plots on intraday timescales for BL Lacertae. The V magnitudes are given on
the X-axis and the (V-R) colour index is plotted against them for each labeled date of observation.}
\label{LC_BL}
\end{figure*}

The STV LC for the target during full monitoring period is displayed in Figure 1.
%The overall magnitude changes in the whole monitoring period are $\Delta B$ = 0.81, $\Delta V$ = 0.74, $\Delta R$ = 0.60 and $\Delta I$ = 0.60.
The short term variability amplitude between JD 2456955.5 and JD 2456984.5 was found to increase with frequency i.e followed the same trend as was
found on intra-night basis with values in B band $\sim$ 81.28\%, V band $\sim$ 74.40\%, R band $\sim$ 59.90\% and I band $\sim$ 59.70\%,
respectively, calculated using equation (4) above.
We have also investigated the corresponding variations in (V-R) colour indices on intraday basis using the analysis criterion described above
and found that the BL Lacertae showed hints of colour variations on 2 of
the nights i.e., on 26 and 27 Oct 2014 with maximum amplitude of variability reaching 6.94\% on 27 Oct.
Behaviour of (V-R) and (B-I) with time as shown in Figure 1 top panel indicates that colour variation is indeed variable on short term basis.
The maximum (V-R) colour variation in the source on short timescale was of 0.838 (between its colour index of 0.826 on JD 2456955.5
and of 0.681 on JD 2456984.5) while for (B-I) maximum
colour variation is found to be 2.726 (between 2.676 on JD 2456955.5 and 2.460 on the last day of our observation span i.e JD 2456984.5.
We got large values of (B-I) as compared to (V-R) as is also evident from Figure 1, which is expected since the standard deviation is likely to
increase with frequency separation between two bands.

\section{\bf Intra- and Inter-band correlations}

To search for the presence of any characteristic timescale of variability we performed the auto-correlation function (ACF) technique discussed in above sections by
auto-correlating the R band which gives the ACF value of $\sim$ 1 with nearly zero time lag and then dropping to negative values
on most of the nights.
From this we inferred that the LCs in R band are auto correlated with themselves and any characteristic variability timescale is absent.

From the intraday LCs in Figure 2, we infer the variations to be simultaneous in both the filters which was confirmed
by DCF analysis technique as explained in section 3.2  with DCF bin sizes from 1 to 10 minutes.
We are getting a strong peak (with DCF $\sim$ 1) at time lag of 0 hrs to 0.007 hrs as clear from the Figure 3.
These time lag values cannot be considered significant since they are close to the measurement intervals.
The small frequency intervals in the optical regime leading to null time lags implies that the photons in these wavebands
are emitting by same physical process and from the same emitting region.
A significant positive lag in V vs R band DCF plots
would imply that the V band variations lead those of the R filter.
The V band magnitude vs R band magnitude plots in Figure 4 also clearly depicts that the flux variations are well correlated on almost all nights
including those when the source was found to be variable indicating that
the variability mechanism seems to be same in both the passbands. A linear model of the form Y = mX + c was used to fit these plots whose results
are given in Table 4 confirming that LCs in both the bands are closely correlated on all those nights when genuine variability was detected.

%For BL Lac, Fan et al.\ (1998) gave an optical variation of $\Delta B$ = 5.3 mag along with the periodicity of $\sim$ 14 years

%As evident from Fig 2 the IDV LCs in different bands are almost consistent with each other.

\section{\bf Colour-magnitude relationship}

We then studied the behaviour of the spectral variations with respect to the brightness in V band for our source.
We fitted straight lines (CI =$m$V + $c$) on each plot of colour index, CI, against V magnitude and fitted values for the slope, $m_2$, along with
constant, $c_2$, are listed in Table 5. Significant positive correlation (the null 
hypothesis probability, $p \le 0.05$) between colour index and apparent blazar magnitude
indicates that the source exhibits a bluer when brighter or redder when fainter trend (H.E.S.S.~Collaboration et al.\ 2014).
In the following analysis, the observed magnitudes have been corrected for galactic extinction based on the extinction map of Schlegel et al.\
(1998) using NED extinction calculator (A$_{V}$ = 0.901 mag, A$_{R}$ = 0.713 mag). The flux from the nucleus of BL Lacertae is contaminated
by the emissions of its giant elliptical host galaxy. So the data is also corrected for the host galaxy contribution, using the measurements of
Nilsson et al.\ (2007) to estimate the host galaxy emission in the R band which is then used to find the corresponding contributions for the
V band (Fukugita et al.\ 1995, Gaur, Gupta, \& Wiita 2012).
During our observations we found that BWB trend
was dominant for the source on intra-night timescales since we are getting significant positive correlation ($p \le 0.05$) between the V magnitude
and (V-R) colour index.
We calculated the slope, $m_2$, and the constant, $c_2$, along with linear Pearson correlation coefficients and the corresponding null 
hypothesis probability, $p_2$ from the colour-magnitude plots (V-R colour Vs V magnitude) for the target for each night, as presented in Figure 6.
On short term basis also we found that (V-R) colour indices follow the BWB trend indicating hardening of the spectrum as the source brightens
with values of $m_3$ = 0.1369 $\pm$ 0.001, $c_3$ = $-$1.2602 $\pm$ 0.01, $r_3$ = 0.4925 $\pm$ 0.002, and $p_3$ = 0.0000.
We need dense, high precision data to clearly know about the trend in short term colour behaviour of the BL Lacertae.

Clear global BWB trend was observed in all 10 nights on intraday timescales with correlation
coefficient ranging between 0.4 and 0.9.
The source brightness was found to have strong BWB chromatism on intraday timescales.
The BWB trend has been found to be predominant in BL Lacertae by all optical observations of this source till date during both flaring and steady states
(Ghosh et al.\ 2000; Gaur et al.\ 2012; Gu et al.\ 2006) which was revealed by our observations also. 
Villata et al.\ (2004b) found that the intraday flares followed BWB trend strongly with a slope of $\sim$ 0.4 which is also the case for our
data with the slopes of the intra-night LCs more than $\sim$ 0.4 in majority of cases. Thus, our colour variability results for the source
being strongly chromatic between Oct - Nov 2014 were consistent with those of Villata et al.\ (2004b).
Since the jet emissions are completely dominating the flux from BL Lacerate objects, this can be explained by
the shock-in-jet models elaborated under section 8.

\begin{table*}
\caption{ Results of linear fit to R Vs V plots and average spectral index values.  }
\textwidth=7.0in
\textheight=10.0in
\vspace*{0.2in}
\noindent

\begin{tabular}{cccccc} \hline 

Observation Date     &  $m_1^a$  &  $c_1^a$  &   $r_1^a$  & $p_1^a$  & $\langle \alpha_{VR} \rangle$  \\ \hline
 26.10.2014          &  0.454 $\pm$ 0.006 & 7.309 $\pm$ 0.081 & 0.862 $\pm$ 0.008  & $<$0.0001  & 4.605 $\pm$ 0.034 \\     
 27.10.2014          &  0.769 $\pm$ 0.004 & 2.630 $\pm$ 0.060 & 0.971 $\pm$ 0.002 & $<$0.0001  & 4.206 $\pm$ 0.038 \\ 
 04.11.2014          &  0.548 $\pm$ 0.016 & 5.805 $\pm$ 0.228 & 0.703 $\pm$ 0.021 & $<$0.0001  & 4.272 $\pm$ 0.046 \\
 05.11.2014          &  0.267 $\pm$ 0.013 & 9.801 $\pm$ 0.183 & 0.388 $\pm$ 0.022 & ~~0.0101  & 4.252 $\pm$ 0.050 \\
 06.11.2014          &  0.885 $\pm$ 0.005 & 0.913 $\pm$ 0.073 & 0.965 $\pm$ 0.002 & $<$0.0001  & 4.297 $\pm$ 0.042 \\
 07.11.2014          &  0.387 $\pm$ 0.017 & 8.324 $\pm$ 0.252 & 0.522 $\pm$ 0.027 & ~~0.0022  & 3.540 $\pm$ 0.047 \\ 
 08.11.2014          &  0.411 $\pm$ 0.015 & 7.890 $\pm$ 0.218 & 0.566 $\pm$ 0.024 & ~~0.0004  & 4.301 $\pm$ 0.045 \\  
 09.11.2014          &  0.302 $\pm$ 0.024 & 9.475 $\pm$ 0.348 & 0.221 $\pm$ 0.011 & ~~0.1310  & 4.240 $\pm$ 0.045 \\  
 16.11.2014          &  0.290 $\pm$ 0.012 & 9.662 $\pm$ 0.176 & 0.409 $\pm$ 0.020 & ~~0.0048  & 3.855 $\pm$ 0.033 \\  
 22.11.2014          &  0.686 $\pm$ 0.014 & 3.768 $\pm$ 0.194 & 0.644 $\pm$ 0.014 & $<$0.0001  & 3.852 $\pm$ 0.030 \\  
\hline
\end{tabular}\\
$^a$ $m_1 =$ slope and $c_1 =$ intercept of CI against V; \\
$r_1 =$ Pearson coefficient; $p_1 =$ null hypothesis probability \\
\end{table*} 

\section{Spectral Energy Distribution (SED)}

Differences in the physical parameters of the relativistic jet causes spectral changes in the emitting electrons which is the reason for SED changes
thus rendering SED studies very useful. It will not only help in understanding the emitting region but also help in discriminating between
models and put tight constraints on those model parameters that are likely to change.

The low energy component of the blazar SED is commonly attributed to the synchrotron process in the relativistic jets while the Compton
up-scattering of the low energy photons is probably the source of high energy component (e.g., Padovani \& Giomini 1995; Sikora \& Madejski 2001;
Coppi 1999). These external photons could be originating from the Accretion Disc by thermal radiation, molecular torus,
reprocessed disc emission from the broad line region (BLR) or from the hot corona (Sikora et al.\ 1994; Dermer \& Schlickeiser 1993). The
accretion disc luminosity along with the distance of the emitting region from the super massive black hole (SMBH) decides the intensity of each external
photon. High energy emission is still not fully understood but for better understanding the proposed theoretical scenarios fall under two
categories i.e the leptonic and the hadronic models. Leptonic models describe the gamma-ray emission produced by IC scattering of photons off the
relativistic electrons while the hadronic models accounts for acceleration of protons in the relativistic Doppler boosted jet to energies sufficient
for photo-pion production which further leads to pion decay and associated particle cascades (Celotti \& Ghisellini 2008; B{\"o}ttcher et al.\ 2007).

In this section, we focus on understanding SED changes associated with the target corresponding to the observed changes in their respective
optical fluxes.
In blazars, optical brightness changes is linked to changes in the spectral shapes and can be quantified by studying the properties of colour
indices.
The calibrated magnitudes of BL Lacertae were de-reddened by subtracting Galactic absorption A$_{B}$ = 1.192 mag, A$_{V}$ = 0.901 mag, A$_{R}$ = 0.713 mag,
and A$_{I}$ =  0.495 mag (Cardelli et al.\ 1989; Bessell et al.\  1998).
Figure 7 shows optical SED for the source,
generated using our B, V, R, I data sets for 13 different epochs with different flux levels.
The faintest SED for BL Lacertae was measured on 2014 Oct 26. Significant variations were
seen in the SEDs during Oct - Nov when it increased in brightness day by day,
reaching a maximum on Nov 22.
%As evident from the figure 7, a hint of the optical bump in the brightest SEDs along with the BWB trend which is common property
%of BL Lacertae objects is clearly visible. 
The synchrotron peak frequency has  values  between 10$^{13}$Hz and 10$^{17}$Hz.
Since the optical band for this source lie near the peak of the synchrotron component of the SED, the optical variability reflects acceleration
processes acting on the highest energy electrons.
Due to the lack of multi-wavelength data here, a detailed description is bit difficult.

\section{Discussion and Conclusion}

To explain optical IDV several theoretical models have been proposed by authors which are broadly classified as extrinsic and intrinsic.
The intrinsic origin of variability in blazars is either due to accretion disc instabilities or relativistic jet activities
and is seen at all accessible timescales ranging from a few tens of minutes through
days and months to even decades. Extrinsic mechanisms include interstellar scintillation
causing radio variations (e.g. Heeschen et al. 1987) at low frequencies and can therefore not be the
case of optical intra-night optical variability (INOV).
Small variations can be explained by turbulence behind an outgoing shock along the jet.
For high SMBH masses (most likely in excess of 10$^8$M$_{\odot}$) the light crossing time is very large which can be linked
with this shortest variability timescales. Among the several accepted theoretical models explaining variability in BL Lacerate objects is the one involving shocks propagating down the in-homogeneous
medium in the relativistic jets. The short timescale variations which we our concerned with like IDV and STV can be attributed to the
irregularities in the jets caused due to the outgoing shocks and high optical polarization (Gaur et al.\ 2014) in case of BL Lacerate objects.
The observed radiation from our source is predominantly non thermal Doppler boosted jet emission (Marscher \& Gear 1985) thus
reducing the observed variability timescale with respect to the rest frame timescale by Doppler factor $\delta$, while increased by (1+z) where z is the
redshift.
Spectrum steepening as magnitude increases can explain the presence of two components, one variable (with a flatter slope,
$\alpha_{1}$) (f$_{\nu}$ $\propto$ $\nu^{-\alpha}$) and the other one stable (with $\alpha_{const}$ $>$ $\alpha_{1}$), contributing to the
overall optical emission. When the variable component dominates we get a chromatic behaviours.
Villata et al.\ (2004a) confirmed that the optical variations on short timescales can be interpreted due to strongly chromatic component while
the longer timescale component as mildly chromatic component. The BWB trend could be explained by one component synchrotron model which tells that more intense the
energy release, the higher the particular frequency (Fiorucci, Ciprini, \& Tostiet al.\ 2004) or
due to the Doppler factor variations in a spectrum deviating
from power law. It may be due to the intrinsic processes related to jet emission mechanism. Luminosity increase due to injection of fresh electrons with the energy distribution harder than that of the previously cooled
ones can also cause flattening of spectrum as the object brightens (Mastichiadis \& Kirk 2002).
According to the shock-in-jet model, as the shock propagating down the jet strikes a region of high electron population, radiations at
different visible colours is produced at different distances behind the shocks. High energy photons from synchrotron mechanism typically
emerge sooner and closer to the shock front than the lower frequency radiation thus causing colour variations. The BWB situation obtained in
this paper will be dominant during the rising phase of a flare which is most common in case of BL Lacerate objects as they are beamed but intrinsically
weak thus causing the observed optical emission from BL Lacerate objects to be overwhelmed by that from the relativistic jets.
Hawkins (2002) said that the effect of the underlying host galaxy and the colour changes from micro-lensing can also lead to colour
variations with time. But since in BL Lacerate objects the Doppler boosting flux from the relativistic jet almost invariably swamps the light from the host galaxy,
the galaxy's light contribution in the observed emission is very less especially during the flaring states, thereby making the red shift
determination very difficult.
Latest studies of Sun et al.\ (2014) discovered a timescale dependent colour variations. They said that the colour of the variable emission
is timescale independent thus ruling out the possibility of BWB trend being due to the contamination from a non-variable redder component, such
as host galaxy. The model states that for timescales $<$ 30 days, the BWB trend in quasars is even stronger and it gradually weakens with
the increasing timescales of above 100 days. Gravitational microlensing is important on weeks to months timescales and is achromatic, so flux and colour
variations observed by us on intraday timescale can be most likely explained by the jet based models explained above.
Since jets in BL Lacerate objects are pointed between $6^{\circ}$ - $10^{\circ}$ with the LOS of the observer, the radiation from it, especially in active
phase is dominated by emissions from the relativistic jets while disc emission is much weaker, hence the jet models
can explain the flux and colour variability detected by us. Doppler boosting greatly amplifies even very weak flux variations produced due to small
changes in the magnetic field or electron density.
\begin{figure}
\epsfig{figure=  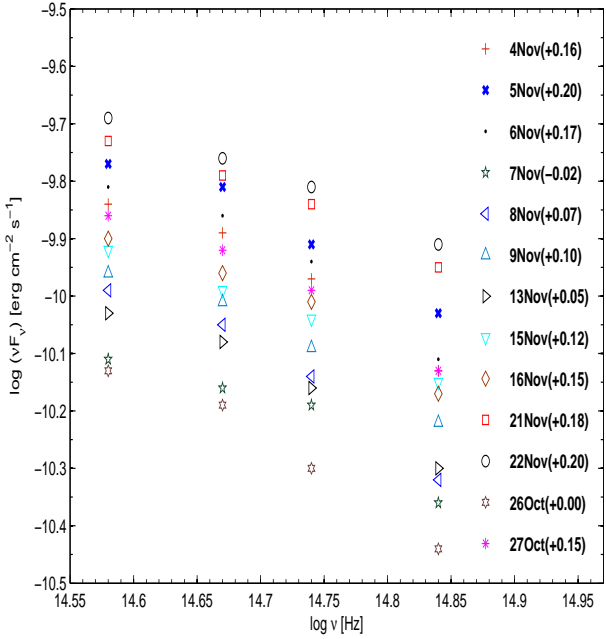,height=3.7in,width=3.6in,angle=0}
\caption{SED results for BL Lacertae in optical frequencies. Different symbols are used for each epoch mentioned in the figure along
with the offset used.}
\label{LC_BL}
\end{figure}
There are several possibilities causing SED changes including freshly injected electrons with energy distribution higher than the average causing
higher synchrotron emission and a peak blue shifting or another reason could be of change in the Doppler boosting factor presumably
due to change in the viewing angle. The changes in the magnetic field can also lead to variations in the SED of BL Lacerate objects. To explore above
mentioned reasons in greater detail we would require data covering the entire EM spectra.

In this paper, we carried out multi-band optical photometry for BL Lacertae between 25 Oct - 23 Nov 2014, in a total of 13 nights, in B, V, R, and
I passbands. Among these 13 nights are 10 dates on which we also carried out quasi-simultaneous observations in V and R filters (as mentioned
in Table 2). We found that the source showed displayed genuine IDV on 3 nights in both V and R bands. If we include PV cases as well then
confirmed microvariability was found on 9 out of 10 nights in R passband while on 5 nights in V filter. 
Variability results can be improved by lengthier observations. Chances of getting variability on intranight timescales are maximum when a blazar
is observed for 6$-$7 hrs as was said by Carini et al.\ (2007) where authors also found that less than 10\% of the sources under study
revealed IDV when observed for $\sim$ 4$-$5 hrs which is the range for most of our observations also.
In the cases where strong variability is found, its amplitude is greater in higher frequency bands which is consistent with Papadakis et al.\
(2003). We also carried out DCF analysis to search for the periodicity and any variability timescale but none was found to be present in
any of the LCs which could be possible if the timescale of electron distribution evolution is longer than the typical monitoring time
(i.e longer than $\sim$ 4 hrs) (Bachev et al.\ 2011). We also searched for correlation
between colour and magnitude and found that our source follows BWB trend which help us to shed some light on the emission mechanism and emission
region. The BL Lacerate objects are on the descending part of the first SED bump where optical spectral index ($\alpha$) should be $>$ 1, which is
also obtained for our source. BL Lacertae was found to have steep optical spectra with spectral index $>$ 3.5 on all observation nights.
We generated spectral energy distribution of the source using the B, V, R, and I data points for all 13 nights with magnitudes in all bands
being de-reddened using the galactic extinction coefficient values. SED studies can provide some useful
information about the possible physical causes of the observed spectral variability.
\begin{table}
\caption{ Fits to colour-magnitude dependencies and colour-magnitude correlation coefficients.  }
\textwidth=7.0in
\textheight=10.0in
\vspace*{0.2in}
\noindent

\begin{tabular}{ccccc} \hline 

Observation Date     &  $m_2^a$  &  $c_2^a$  &   $r_2^a$  & $p_2^a$  \\ \hline

26.10.2014 & 0.546 $\pm$ 0.006 & $-$7.031 $\pm$ 0.077 & 0.898 $\pm$ 0.006 & $<$0.0001  \\
27.10.2014 & 0.229 $\pm$ 0.004 & $-$2.608 $\pm$ 0.056 & 0.771 $\pm$ 0.013 & $<$0.0001  \\
04.11.2014 & 0.450 $\pm$ 0.016 & $-$5.590 $\pm$ 0.214 & 0.630 $\pm$ 0.025 & $<$0.0001  \\
05.11.2014 & 0.733 $\pm$ 0.013 & $-$9.368 $\pm$ 0.172 & 0.756 $\pm$ 0.013 & $<$0.0001 \\
06.11.2014 & 0.115 $\pm$ 0.005 & $-$1.000 $\pm$ 0.069 & 0.431 $\pm$ 0.022 & ~~0.0049 \\
07.11.2014 & 0.611 $\pm$ 0.017 & $-$7.966 $\pm$ 0.239 & 0.693 $\pm$ 0.021 & $<$0.0001 \\
08.11.2014 & 0.589 $\pm$ 0.015 & $-$7.577 $\pm$ 0.206 & 0.701 $\pm$ 0.019 & $<$0.0001 \\
09.11.2014 & 0.698 $\pm$ 0.024 & $-$9.077 $\pm$ 0.328 & 0.464 $\pm$ 0.018 & ~~0.0009 \\
16.11.2014 & 0.710 $\pm$ 0.012 & $-$9.258 $\pm$ 0.167 & 0.731 $\pm$ 0.012 & $<$0.0001 \\
22.11.2014 & 0.312 $\pm$ 0.014 & $-$3.670 $\pm$ 0.182 & 0.356 $\pm$ 0.019 & ~~0.0104 \\
\hline
\end{tabular}  \\
$^a$ $m_2 =$ slope and $c_2 =$ intercept of CI against V; \\
$r_2 =$ Pearson coefficient; $p_2 =$ null hypothesis probability \\
\end{table} 

\section*{Acknowledgments}
We thank the referee for detailed and thoughtful comments which helped us to improve the manuscript.

\label{lastpage}
\end{document}